\documentclass{llncs}

\usepackage{versions}

\excludeversion{SPW14}
\includeversion{LONG}
\usepackage[english]{babel}  
\usepackage[utf8]{inputenc} 
\usepackage[T1]{fontenc} 
\usepackage{indentfirst}
\usepackage{multicol}
\usepackage{amsmath}
\usepackage{amssymb}
\usepackage{booktabs}
\usepackage{hyperref}
\usepackage{subfig}
\usepackage{soul}
\usepackage{bm}
\usepackage{bbding}
\usepackage{enumitem} 
\usepackage{array}
\usepackage{caption}
\usepackage{fancyhdr}
\usepackage{proof}
\usepackage{ifsym}
\usepackage{color}
\usepackage{url}

\newcommand{\mathii}[1] {\emph{#1}}
\newcommand{\mathbi}[1]{\textbf{\em #1}}

\newcommand{\Agents}{\mathit{Agents}}
\newcommand{\Eves}{\mathit{Dishonest}}
\newcommand{\Honest}{\mathit{Honest}}
\newcommand{\bEves}{\mathit{BenignDishonest}}

\newcommand{\DistinguisherC}{\Delta_C}
\newcommand{\DistinguisherR}{\Delta_\mathit{Id}}

\newcommand{\dState}{\delta}
\newcommand{\Invariant}{\mathit{Inv}}
\newcommand{\iState}{\iota}
\newcommand{\fState}{\phi}
\newcommand{\rState}{\lambda}
\newcommand{\infState}{\rho}

\newcommand{\puntello}{\rule[-2.5 mm]{0mm}{0.65 cm}}

\newcommand{\canseeM}[1]{\mathit{canSee}(#1))}
\newcommand{\ofinterest}[1]{$\mathit{ofInterest_{E}(#1)}$}
\newcommand{\canseeNoArg}{$\mathit{canSee}$}
\newcommand{\nethandler}{$\mathit{NetHandler}$}
\newcommand{\trueid}{$\mathit{True\text{-}Sender \text{-}ID}$}

\newcommand{\triple}{$\langle \mathit{sender\textrm{-}ID, message, receiver\textrm{-}ID}\rangle$}

\newcolumntype{L}[1]{>{\raggedright\let\newline\\\arraybackslash\hspace{0pt}}m{#1}}
\newcolumntype{C}[1]{>{\centering\let\newline\\\arraybackslash\hspace{0pt}}m{#1}}
\newcolumntype{R}[1]{>{\raggedleft\let\newline\\\arraybackslash\hspace{0pt}}m{#1}}

\newcommand{\fix}[2]{{\bf FIX}\footnote{{\bf #1:} #2 }}

\usepackage{tikz}
\usetikzlibrary{shadows,matrix,petri,patterns,fit,backgrounds,automata,arrows,positioning,shapes}

\DeclareCaptionType{copyrightbox}

\urldef{\mailsa}\path|{michele.peroli,luca.vigano,matteo.zavatteri}@univr.it|

\begin{SPW14}
\title{Non-collaborative Attackers and How and Where to Defend Flawed Security Protocols\thanks{This work was partially supported by the EU
	FP7 Project no. 257876, ``SPaCIoS: Secure Provision and Consumption in
	the Internet of Services'' (www.spacios.eu) and the PRIN 2010-11 project ``Security Horizons''. Much of this work was carried out while Luca Vigan\`o was at the Universit\`a di Verona.} 
}
\end{SPW14}

\begin{LONG}
\title{Non-collaborative Attackers and How and Where to Defend Flawed Security Protocols \\
(Extended Version)\thanks{This work was partially supported by the
EU FP7 Project no. 257876, ``SPaCIoS: Secure Provision and
Consumption in the Internet of Services'' (www.spacios.eu) and the
PRIN 2010-11 project ``Security Horizons''. Much of this work was
carried out while Luca Vigan\`o was at the Universit\`a di Verona.}
}
\end{LONG}

\author{Michele Peroli\inst{1} \and Luca Vigan\`o\inst{2} \and Matteo Zavatteri\inst{1}}

\institute{Dipartimento di Informatica, Universit\`a di Verona, Italy 
\and
Department of Informatics, King's College London, UK}

\begin{document}

\maketitle 
\begin{abstract}
Security protocols are often found to be flawed after their
deployment. We present an approach that aims at the neutralization or
mitigation of the attacks to flawed protocols: it avoids the
complete dismissal of the interested protocol and allows honest agents
to continue to use it until a corrected version is released. Our
approach is based on the knowledge of the network topology, which we
model as a graph, and on the consequent possibility of creating an
interference to an ongoing attack of a Dolev-Yao attacker, by means of
non-collaboration actuated by ad-hoc benign attackers that play the role
of network guardians. Such guardians, positioned in strategical points
of the network, have the task of monitoring the messages in transit and
discovering at runtime, through particular types of inference, whether
an attack is ongoing, interrupting the run of the protocol in the
positive case. We study not only how but also where we can attempt to
defend flawed security protocols: we investigate the different
network topologies that make security protocol defense feasible and
illustrate our approach by means of concrete examples.
\end{abstract}


\section{Introduction}

\subsection{Context and motivation} 
Security protocols are often found to be flawed after their
deployment, which typically requires ``dismissing'' the protocol
and hurrying up with the deployment of a new version hoping to be
faster than those attempting to exploit the discovered flaw. We
present an approach that aims at the neutralization or mitigation
of the attacks to flawed protocols: it avoids the complete
dismissal of the interested protocol and gives honest agents the
chance to continue to use it until a corrected version is released.

The standard attacker model adopted in security protocol analysis
is the one of~\cite{DY:1983}: the \emph{Dolev-Yao (DY) attacker}
can compose, send and intercept messages at will, but, following
the perfect cryptography assumption, he cannot break cryptography.
The DY attacker is thus in complete control of the network --- in
fact, he is often formalized as being the network itself --- and,
with respect to network abilities, he is actually stronger than any
attacker that can be implemented in real-life situations. Hence, if
a protocol is proved to be secure under the DY attacker, it will
also withstand attacks carried out by less powerful attackers;
aside from deviations from the specification (and the consequent
possible novel flaws) introduced in the implementation phase, the
protocol can thus be safely employed in real-life networks, at
least in principle.

A number of tools have been proposed for automated security
protocol analysis (e.g.,
\cite{AVANTSSAR,proverif,Cr2008Scyther,EscobarMM07,Ryan00,avispa02}
to name just a few), all of which follow the classical approach for
security protocol analysis in which there is a finite number of
honest agents and only one DY dishonest agent, given the implicit
assumption that in order to find attacks we can reduce $n$
collaborative DY attackers to $1$ (for a proof of this assumption
see, e.g., \cite{Basin:2011:DTL:1994484.1995278}).

In this paper, we take a quite different approach: we exploit the
fact that if in the network there are \emph{multiple
non-collaborative attackers}, then the interactions between them
make it impossible to reduce their attack ``power'' to that of a
single attacker. This paper is based on the network suitable for
the study of non-collaborative scenarios defined in our previous
works~\cite{FPV:02SECRYPTbook,FPV:SECOTS}, in which we introduced a
protocol-independent model for non-collaboration for the analysis
of security protocols (inspired by the exploratory
works~\cite{Bella03,retaliation} for ``protocol life after
attacks'' and attack retaliation). In this model: (i) a protocol is
run in the presence of multiple attackers, and (ii) attackers
potentially have different capabilities, different knowledge and do
not collaborate but rather may interfere with each other.

Interference between attackers has spawned the definition of an ad
hoc attacker, called \emph{guardian}, as a defense mechanism for
flawed protocols: if two non-collaborative attackers can interfere
with each other, then we can exploit this interference to
neutralize or at least mitigate an ongoing attack (a detailed
cost-effective analysis of this approach is left for future
work).\footnote{It is interesting to note how this idea of ``living
with flaws'' is becoming more and more widespread; see, e.g.,
\cite{mithys} where runtime monitors are employed to warn users of
android applications about ``man in the middle'' attacks on flawed
implementations of SSL. Our approach is also related to
signature-based intrusion detection systems, but we leave the
detailed study of the relations of our approach with runtime
monitors and signature-based intrusion detection systems for future
work.}

There is one fundamental catch, though. We know that a DY attacker
actually cannot exist (e.g., how could he control the whole
network?) but postulating his existence allows us to consider the
worst case analysis so that if we can prove a protocol secure under
such an attacker, then we are guaranteed that the protocol will be
secure also in the presence of weaker, more realistic attackers. A
guardian, however, only makes sense if it really exists, i.e., if
it is implemented to defend flawed protocols for real, but the
attackers and the guardian presented
in~\cite{FPV:02SECRYPTbook,FPV:SECOTS} are modeled in order to
discover interactions between agents in non-collaborative scenarios
rather than pushing for an implementation in the real-world.

\subsection{Contributions}
Since implementing a guardian with the full power of a DY attacker
is impossible, we must investigate ways to make the guardian more
feasible. In order to reduce the complexity of the possible
implementation of such a defense mechanism, in this paper we relax
the notion of guardian and ask him to defend only a subset of the
communication channels of the network, which we put under his
control.

Furthermore, not being obviously able to know where the competitor
is, we investigate where we have to introduce this defense
mechanism in the network from a topological perspective, i.e., how
the guardian can dominate his competitor(s).\footnote{In the
following, we focus on one competitor (i.e., one attacker), but it
is quite straightforwardly possible to extend our work to multiple
competitors.} Modeling the network as a graph, we study how the
topological position of an attacker $E$ and a guardian $G$, with
respect to each other and to honest agents of the protocol, can
influence a protocol attack and, thus, the possible defense against
it. We define six basic topological configurations and study the
outcome of the introduction of a guardian in each specific
position. We also introduce the concept of \emph{topological
advantage}, which guarantees that the guardian has an advantage
with respect to his competitors, and can thus carry out inference
on messages in transit in order to detect an ongoing attack and
eventually mitigate or neutralize it.

The contributions of this paper thus extend, and in a sense are
complementary to, the ones in our previous
works~\cite{FPV:02SECRYPTbook,FPV:SECOTS}. In a nutshell: there we
discussed the \emph{how} we can defend flawed security protocols
and here we discuss the \emph{where}. More specifically, as we will
describe in the following sections,
in~\cite{FPV:02SECRYPTbook,FPV:SECOTS}, we put the basis for the
study of the interaction of two attackers in non-collaborative
scenarios with the goal of understanding and finding the types of
interference the guardian can use, and, in this paper, we give the
means to understand how to exploit the interference from a
topological point of view, thus bringing the guardian close to real
implementation.

\subsection{Organization}

We proceed as follows. In Section~\ref{sec:interference}, we
summarize the main notions of attack interference in
non-collaborative scenarios. In Section~\ref{sec:modeling}, we
formalize the models of the network and of the guardian, with
particular emphasis on the topological advantage that a guardian
must have in order to defend against attacks. 
\begin{SPW14}
In Section~\ref{sec:case-studies}, we discuss, as a detailed
proof-of-concept, how we can defend the ISO-SC 27 protocol and
summarize the results we obtained for other case studies.
\end{SPW14}
\begin{LONG}
In Section~\ref{sec:case-studies}, we discuss, as a detailed
proof-of-concept, how we can defend the ISO-SC 27 protocol and
summarize the results we obtained for other case studies, which are
described in more detail in the appendix.
\end{LONG}
In Section~\ref{sec:conclusions}, we briefly summarize our results and discuss future work.

\section{Attack interference in non-collaborative networks}
\label{sec:interference}

\subsection{Network agents} 

Let $\Agents$ be the set of all the network agents, which comprises
of two disjoint subsets: \begin{itemize} \item the subset $\Honest$
of \emph{honest agents} who always follow the steps of the security
protocol they are executing in the hope of achieving the properties
for which the protocol has been designed (such as authentication
and secrecy), and \item the subset $\Eves$ of \emph{dishonest
agents} (a.k.a. \emph{attackers}) who may eventually not follow the
protocol to attack some (or all) security properties. In addition
to being able to act as legitimate agents of the network, dishonest
agents typically have far more capabilities than honest agents and
follow the model of Dolev-Yao~\cite{DY:1983} that we summarized in
the introduction. \end{itemize}

The \emph{knowledge} of an honest agent $X$ is characterized by a
proprietary dataset $D_X$, which contains all the information that
$X$ acquired during the protocol execution, and is closed under all
cryptographic operations on message terms (e.g., an agent can
decrypt an encrypted message that he knows provided that he knows
also the corresponding decryption key). $D_X$ is monotonic since an
agent does not forget.

\subsection{DY attackers and the network in a non-collaborative scenario}

	\begin{table}[!t]
	\caption[Attacker model for non-collaboration]{Dolev-Yao attacker model for non-collaborative scenarios: internal operations (synthesis and analysis of messages), network operations (\mathii{spy}, \mathii{inject}, \mathii{erase}) and system configuration (\trueid{}, \mathii{DecisionalProcess}, \nethandler{}). \nethandler{} describes the set of attackers who are allowed to spy by applying one of the \mathii{spy} rules. We omit the usual rules for conjunction.
	\label{OurDYmodel}
	}	
	\begin{center}
	\scalebox{0.82}{
	\begin{tabular}{c}
	\begin{tabular}{cccc}
	\hline
	\hline \\
	\mathbi{Composition:} &
	\mathbi{Encryption: \phantom{\quad}} &
	\mathbi{Projection:\phantom{\quad}} &
	\mathbi{Decryption:}
	\\\\
	$\infer[]{(m_{1}, m_{2})\in D^{i}_{E}}{m_{1}\in D^{i}_{E} \quad m_{2}\in D^{i}_{E}}$\quad ~
	&
	$\infer[]{\{m\}_{k}\in D^{i}_{E}}{m\in D^{i}_{E} \quad k\in D^{i}_{E}}$
	\quad ~
%	\\\\
	&
	$\infer[]{m_{j}\in D^{i}_{E}  \:\: \text{for~} j \in \{1,2\}}{(m_{1}, m_{2})\in D^{i}_{E}}$ \quad ~
	&
	$\infer[]{m \in D^{i}_{E}}{\{m\}_{k}\in D^{i}_{E} \quad k^{-1} \in D^{i}_{E}}$
	\end{tabular}~
	\\\\
	\hline
	\hline \\
	% \mathbi{Restricted-Spy}:
	% \\~\\
	% $\infer[]{m \in D^{i+1}_{E}}{<X, m, Y> \in D_{net}^{i} \quad \text{sender}(<X,m,Y>) \in D^{i}_{E} \quad Y \in D^{i}_{E}\quad \psi}$ 
	% \\\\
\begin{tabular}{cc}
	\mathbi{Inflow-Spy}: 
	&
	\mathbi{Outflow-Spy}:
	\\~\\
	$\infer[]{m \in D^{i+1}_{E} \land sender(<X,m,Y>) \in D^{i+1}_{E} }{\mu \in D_{net}^{i} \quad $ \ofinterest{X}$ \quad Y \in D^{i}_{E} \quad \psi} $
	\qquad ~
&
	$\infer[]{m \in D^{i+1}_{E} \land Y \in D^{i+1}_{E}}{\mu \in D_{net}^{i} 	\quad {sender}(\mu) \in D^{i}_{E} \quad $\ofinterest{Y}$ \quad \psi }$ 
\end{tabular}
	\\\\
	{where $\mu = <X, m, Y>$ \quad and \quad $\psi = E \in \canseeM{<\!X,m,Y\!>, i}$  }\\\\
	\hline
	\\
	\begin{tabular}{cc}
		\mathbi{Injection}: 
		&
		\mathbi{Erase}: 
		\\~\\
		$\infer[]{<E(X), m, Y>  \in D_{net}^{i+1}}{m \in D^{i}_{E} \quad X \in D^{i}_{E} \quad Y \in D^{i}_{E}}$ 
		\qquad ~
	&
	$\infer[]{<X, m, Y> \notin D_{net}^{i+1}}{<X,m,Y> \in D_{net}^{i} \quad {sender}(<X,m,Y>) \in D^{i}_{E}}$
	\end{tabular}
	\\\\
	\hline \hline
	\\ 
	\mathbi{True-sender-ID}:
	\\~\\
	$
	{sender}(<X, m, Y>)= \begin{cases}
	 E & \text{~if~ there exists $Z$ such that~} X = E(Z) \\
	 X & \text{~otherwise}
	\end{cases}
	$  
	\\~\\
	\mathbi{DecisionalProcess}:
	\\~\\
	\ofinterest{X}
	$
	=  
	\begin{cases}
	 true & \text{if $E$ decides to pay attention to $X$} \\
	 false & \text{~otherwise}
	\end{cases}
	$
	%{\scriptsize $(D)$}
	 \\\\
	\hline
	\\
	\mathbi{NetHandler}:
	\\~\\
	$
	\canseeM{<\!X,m,Y\!>,i} \: = \:%&
	 \{Z \in \Eves \:|\:   \text{ $Z$ can spy}%\\
	% & 
	 <X,m,Y> \text{~on $D_{net}^{i}$} \}
	$
	\\\\
	\hline
	\hline
	\end{tabular}
	}
	\end{center}
	\end{table}

In this paper, we take the non-classical approach that leverages on the
fact that the interactions between multiple non-collaborative attackers
may lead to interference. We base our work on the network suitable for
the study of non-collaborative scenarios defined
in~\cite{FPV:02SECRYPTbook,FPV:SECOTS}, which we now summarize quickly
pointing to these two papers for more details.

Table~\ref{OurDYmodel} shows the model that we adopt to formalize a
DY attacker $E$ in a non-collaborative scenario in which different
attacks may interfere with each other (we restrict the study of
this type of interaction to two active attackers but it can be
generalized to multiple ones). The knowledge base of $E$ is encoded
in the set $D_{E}$, whereas $D_{net}$ is the proprietary dataset
for the network (we will return to the network model below). The
rules in the table describe the operations that an attacker can
perform internally, how he can interact with the network and how
the system (i.e., the network environment) is configured. It is
important to note that the rules in Table~\ref{OurDYmodel} are
transition rules rather than deduction rules, i.e., they describe
knowledge acquisition from a given operation and a particular
configuration rather than the reasoning about ``only'' the
knowledge of the attacker.

As in the classic DY case, an attacker in this model can
\emph{send} and \emph{receive} messages, derive new messages by
\emph{composing, decomposing, modifying, encrypting/decrypting}
known messages (iff he has the right keys), and \emph{intercept} or
\emph{remove} messages from the network. An attacker $E$ may also
masquerade as (i.e., impersonate) another agent $X$, which we
denote by writing $E(X)$.

The most significant features of the attacker abilities are the two
\emph{spy} rules, which formalize the fact that attackers only pay
attention to a selection of the traffic on the network (considering only selected target agents):\footnote{If an attacker were omniscient and omnipotent (i.e., if he were to control the whole network) then there'd actually be no ``space'' for another attacker, and thus there'd be no interference. The more ``adventurous'' reader may want to compare this with the proof of the uniqueness of God by Leibniz, which was based on the arguments started by Anselm of Canterbury and was later further refined by G\"odel.}
\footnote{In this paper
we only use the \emph{inflow-spy} and the \emph{outflow-spy} filters
and not the \emph{restricted-spy} filter used in the previous exploratory
works. This is due to the fact that we can certainly know who we want to
defend, but we cannot know who the attackers are and we want to have
the possibility of intercepting \emph{all} outgoing/incoming messages
which leave/come from/to an agent $X$.}

\begin{itemize} 
\item {\small $\mathbi{Inflow-Spy}$}: the attacker pays attention to the
incoming network traffic of a target agent and saves the identifiers of
the sender agents,
\item {\small $\mathbi{Outflow-Spy}$}: the attacker pays attention to
the traffic generated by a target agent.
\end{itemize}

The \emph{target agent} $X$ of the two spy-rules is defined through a
decisional process (the function \ofinterest{X} in
Table~\ref{OurDYmodel}) in which each attacker decides if the traffic
to/from the agent $X$ is worth to be followed. This decision is made at
run-time when a new agent identifier is discovered over the network
(i.e., when a new agent starts sending messages on the channel monitored
by the attacker). In this paper, we do not go into the details of how his decision is actually taken, but different strategies might be devised and we will investigate them in future work.

The network net is also formalized through a dataset, $D_{net}$, which
is changed by the \emph{actions} send, receive, inject and erase a
message. We write $D_{net}^i$ to denote the state of $D_{net}$ after the
$i$-th action. Messages transit on the network in the form of triplets
of the type 
\begin{displaymath}
\textit{\triple{}}, 
\end{displaymath}
where, as in the classical approaches, both the
attackers and the agents acquire knowledge only from the body of
messages, i.e., $\mathit{sender\textrm{-}ID}$ and
$\mathit{receiver\textrm{-}ID}$ are actually hidden to them and only
used by the network system. As a consequence of message delivery or
deletion, $D_{net}$ is non-monotonic by construction.

In order to regulate the concurrent actions over the network, the model
comprises a \nethandler{} whose task is to handle the network by
selecting the next action and implementing the dependencies between
selected actions and knowledge available to each attacker. That is,
\nethandler{}: (i) notifies agents that the state of the network has
changed with newly-inserted messages, (ii) polls agents for their next
intended action, (iii) selects from the set of candidate actions the one
that will be actually carried out, and (iv) informs agents of whether
the computation they performed to propose an action is a consequence of
a message that they did not have access to (i.e., for these agents a
rollback might occur in which all knowledge gained since the last
confirmed action is deleted from the dataset, and internal operations
that have occurred are cancelled).

The outcome of the process governed by the network handler is described
through the function \canseeNoArg{}, which returns a subset of dishonest
agents, highlighting the identifier of attackers who can spy ``before''
the message is erased from $D_{net}$. In other words, when a message is
deleted from the network, the network handler, through the function
\canseeNoArg{}, can decide if an attacker has spied (and saved in his
dataset) the message or not.
%\fix{M}{modified sentence}
In our previous work we had the possibility of spying a message before its 
deletion (in this case, the attacker has to decide if the message has been 
received by the honest agent or deleted by another attacker) but in this paper 
we relax this assumption and decide that when a message is spied it remains in 
the dataset of the attacker. 
The function \canseeNoArg{} is a configurable parameter of
our network and it corresponds to configuring a particular network
environment in which the agents are immersed: \canseeNoArg{} is instantiated 
by the security analyst at the beginning of the analysis in order to model 
time-dependent accessibility, strategic decision-making and information-sharing, 
or to capture a particular network topology (in our framework the function \canseeNoArg{} is necessary in order to model the topologies that we will introduce in  Section~\ref{sec:networkTA}).

\subsection{Attack interference (in the case of the ISO-SC 27 protocol)}

\begin{table}[t!]
	\caption{The ISO-SC 27 protocol and a parallel session attack against it.\label{tab:isoSC27}}
\begin{center}
\scalebox{1}{
\begin{tabular}{cc}
  \begin{tabular}{c}	\hline
{\bf ISO-SC 27 protocol} \\
\hline 
\\
 $
\begin{array}{rll}
	(1) & A \to B  & : N_A \\
	(2) & B \to A  & : \{\!| N_A, N_B |\!\}_{K_{AB}}\\
	(3) & A \to B  & : N_B \\
\end{array}
 $
\\ \\
\hline
\end{tabular}
&
$~ \qquad \qquad ~$
\begin{tabular}{c}
 \hline
	{\bf Attack}\\
 \hline
\\
 $
\begin{array}{rll}
	(1.1) & A 	 \to E(B) & : N_A \\
	(2.1) & E(B) \to A    & : N_A \\
	(2.2) & A 	 \to E(B) & : \{\!| N_A, N'_A |\!\}_{K_{AB}}\\
	(1.2) & E(B) \to A    & : \{\!| N_A, N'_A |\!\}_{K_{AB}}\\
	(1.3) & A 	 \to E(B) & : N'_A \\
	(2.3) & E(B) \to A    & : N'_A \\
\end{array}
 $
\\\\
\hline
\end{tabular}
\end{tabular}
}
\end{center}
\end{table}

As a concrete, albeit simple, example of security protocol,
Table~\ref{tab:isoSC27} shows the ISO-SC 27 protocol \cite{ISOSC27}, which 
aims to achieve entity authentication (aliveness) between two honest agents 
$A$ and $B$, by exchanging nonces, under the assumption that they already 
share a symmetric key $K_{AB}$. 
Since in the second message there
is nothing that assures that the message actually comes from $B$, the
protocol is subject to a parallel sessions attack (also shown in the table) in which the attacker
$E$, who does not know $K_{AB}$, uses $A$ as oracle against herself in
order to provide to her a response that he cannot generate by himself:
$E$ masquerades as $B$ intercepting $A$'s first message and sending it
back to her in a parallel session (messages (1.1) and (2.1)). When $A$
receives the first message of the protocol from $E$, she thinks someone
wants to talk with her in another instance of the protocol (she does not
control the nonce), thus she replies to $E$ generating another nonce $N_A'$  and encrypting it 
together with $N_A$ (message (2.2)). 
Now $E$ has got everything he needs in order to complete the attack to the
protocol (messages (1.2)). The last message is not mandatory as
the session has already been attacked, thus $E$ can omit it (message
(2.3)). At the end of the protocol runs, $A$ is fooled into believing
that $E(B)$ is $B$.

If a protocol is flawed, a single DY attacker will succeed with
certainty. However, if attacks to the same protocol are carried out in a
more complex network environment, then success is not guaranteed since
multiple non-collaborative attackers may interact, and actually
interfere, with each other. The results
of~\cite{FPV:02SECRYPTbook,FPV:SECOTS} show that it is possible, at
least theoretically, to exploit interference between two
non-collaborative attackers to mitigate protocol flaws, thus providing a
form of defense to flawed protocols.

In the case of ISO-SC 27 protocol, which was not studied
in~\cite{FPV:02SECRYPTbook,FPV:SECOTS}\footnote{In~\cite{FPV:02SECRYPTbook,FPV:SECOTS}, we analyzed two protocols: (i) a
key transport protocol described as an example in~\cite{boydMathuria},
which we thus called the Boyd-Mathuria Example (BME), and (ii) the
Shamir-Rivest-Adleman Three-Pass protocol
(SRA3P~\cite{Clark97:asurvey}), which has been proposed to transmit data
securely on insecure channels, bypassing the difficulties connected to
the absence of prior agreements between the agents $A$ and $B$ to
establish a shared key.
},
we can identify six cases for
the possible interaction between two non-collaborative attackers $E_1$
and $E_2$:
\begin{enumerate}\itemsep0em
	\item $E_{1}$ and $E_{2}$ know each other as honest.
	\item $E_{1}$ and $E_{2}$ know each other as attackers.
	\item $E_{1}$ and $E_{2}$ are unaware of each other.
	\item $E_{2}$ knows $E_{1}$ as honest.
	\item $E_{2}$ knows $E_{1}$ as dishonest.
	\item $E_{2}$ knows $E_{1}$, but he is unsure of $E_{1}$'s honesty.
\end{enumerate}
The traces corresponding to the interactions of $E_1$ and $E_2$
attacking the protocol are shown in Table~\ref{tracesBME}. Attack
traces of this type lead to three possible (mutually exclusive)
situations: (i) $E_1$ dominates $E_2$ (i.e., $E_1$'s attack
succeeds while $E_2$'s fails), or (ii) none of their attacks has
success, or (iii) both achieve a situation of uncertainty, i.e.,
they do not know if their attacks have been successful or not.

In order to exploit the interference generated by multiple dishonest
agents attacking the same protocol, we can construct an additional, but
this time non-malicious, attacker: the \emph{guardian} $G$. 

To define the guardian as a network agent, we refine the previous
definition of $\Agents$ to consider the subset of \emph{benign dishonest
agents}, i.e., $\bEves \subseteq \Eves \subseteq \Agents$, where $X \in
\bEves$ means that $X$ has attacker capabilities and may not follow the
protocol but he ``attacks'' with the goal of ``defending'' the security
properties not of attacking them.
In other words: 
\begin{definition}[Guardian]
A \emph{guardian} is a benign dishonest agent of the network,
transparent to the other agents, whose main task is to establish a
partial (or total) defense mechanism in order to mitigate (or
neutralize) protocol attacks at execution time by means of
attack-interference in non-collaborative scenarios.
$G$ is transparent to honest agents during their execution and becomes ``visible'' only in the case he has to report an ongoing attack. 
\end{definition}

\begin{table}[t!]
	\caption{Traces for non-collaborative attacks against the ISO-SC 27. Traces are exhaustive: $E_1$ and $E_2$ have priority over honest agents. Arrows: relative order between $(2.1')$ and $(2.1'')$ is irrelevant in determining the outcome.}
	\label{tracesBME}
	% \uparrow
\begin{center}
\scalebox{1}{
\begin{tabular}{cc}
 \hline
 \footnotesize {\bf  T1: cases 1, 3, 4} 
& 
\footnotesize {\bf  T2: cases 5}\\
\hline \\
\footnotesize
$
\begin{array}{rll}
(1.1) & A \to E_{1,2}(B)  & : N_A \\
(2.1) & E_{1,2}(B) \to A  & : N_A \\
(2.2) & A \to E_{1,2}(B)  & : \{\!| N_{A}, N'_A |\!\}_{K_{AB}} \\
(1.2) & E_{1,2}(B) \to A  & : \{\!| N_{A}, N'_A |\!\}_{K_{AB}} \\
(1.3) & A \to E_{1,2}(B)  & :  N'_{A} \\
(2.3) & E_{1,2}(B) \to A  & :  N'_{A} 
\end{array}
$
& \qquad
\footnotesize
$ 
\begin{array}{rll}
(1.1\phantom{''}) & A \to E_{1,2}(B)  & : N_A \\
\downarrow(2.1'\phantom{'}) & E_{1}(B) \to E_{2}(A)  & : N_A \\
\uparrow(2.1'') & E_{2}(B) \to A  & : N_A \\
(2.2\phantom{''}) & A \to E_{2}(B)  & : \{\!| N_{A}, N'_A |\!\}_{K_{AB}} \\
(1.2\phantom{''}) & E_{2}(B) \to A  & : \{\!| N_{A}, N'_A |\!\}_{K_{AB}} \\
(1.3\phantom{''}) & A \to E_{2}(B)  & :  N'_{A} \\
(2.3\phantom{''}) & E_{2}(B) \to A  & :  N'_{A} 
\end{array}
$
\\ 
&\\
\hline
\footnotesize {\bf T3: case 2} & 
\footnotesize {\bf T4: case 6} \\
\hline \\
\footnotesize
$
\begin{array}{rll}
	(1.1\phantom{''}) & A \to E_{1,2}(B)  & : N_A \\
	\downarrow(2.1'\phantom{'}) & E_{1}(B) \to E_{2}(A)  & : N_A \\
	\uparrow(2.1'') & E_{2}(B) \to E_{1}(A)  & : N_A \\
\end{array}
$
&
\footnotesize
$
\begin{array}{rll}
	(1.1) & A \to E_{1,2}(B)  & : N_A \\
	(2.1) & E_{1}(B) \to A  & : N_A \\
	\multicolumn{3}{c}{\text{ + steps of case 5}} \\
\end{array}
$
\\ 
&\\
\hline
\hline
\end{tabular}
}
\end{center}
\end{table}

\section{Modeling the network and the guardian}
\label{sec:modeling}

In the previous section, we have seen how the interaction between
multiple non-collaborative dishonest agents attacking the same protocol
can interfere with both attacks, thus providing a form of defense. 
As we remarked in the introduction, even if the idea of having a
guardian defending honest agents from attacks seems thrilling, the existence
of a guardian agent makes sense only with his implementation in the real
world. In order to reduce the complexity of such an implementation, we
will now investigate where we have to introduce this defense
mechanism in the network from a topological perspective (i.e., how the
guardian can dominate his competitor(s)). Modeling the network as a graph,
we study how the topological position of an attacker $E$ and a guardian
$G$, with respect to each other and to honest agents of the protocol,
can influence a protocol attack and, thus, the possible defense against
it.

We say that the outcome of the introduction of the guardian on the
network for a particular protocol yields a:
\begin{itemize}
	\item \emph{false positive} if, for some reason, a normal run of the protocol is considered as an attack,
	\item \emph{false negative} if, for some reason, an attack is considered as a normal run of the protocol,
	\item \emph{partial defense} iff it admits false negatives,
	\item \emph{total defense} iff it does not admit false negatives.
\end{itemize}

Our objective is to realize a defense mechanism that admits as few false
negatives as possible, while limiting also the number of false positives, by investigating the position that gives the guardian a topological advantage (see  Definition~\ref{def:def-mechanism} of defense mechanism and the ensuing Theorem~\ref{thm:defense}).

\subsection{A network for topological advantage}
\label{sec:networkTA}
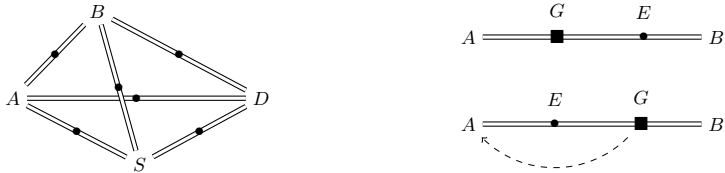
\begin{figure*}[!t]
\centering
\subfloat[][\footnotesize An example of network as a graph; vertices represent agents, edges represent communication channels and the bullets $\bullet$ represent the presence of a DY-attacker $E$.]
{
\parbox{0.4\textwidth}
	{\centering
	\scalebox{0.8}{
	\begin{tikzpicture}[font=\small]
		\node (A) {$A$};
		\node (B) [above=of A,xshift=40pt] {$B$};
		\node (S) [below=of A,xshift=60pt,yshift=10pt] {$S$};
		\node (D) [right=of A,xshift=75pt] {$D$};
	
		\draw[double,double distance=1.5pt,line width=0.5pt] (A) to node {$\bullet$} (B);
		\draw[double,double distance=1.5pt,line width=0.5pt] (A) to node {$\bullet$} (S);
		\draw[double,double distance=1.5pt,line width=0.5pt] (A) to node {$\bullet$} (D);
		\draw[double,double distance=1.5pt,line width=0.5pt] (B) to node {$\bullet$} (S);
		\draw[double,double distance=1.5pt,line width=0.5pt] (B) to node {$\bullet$} (D);
		\draw[double,double distance=1.5pt,line width=0.5pt] (S) to node {$\bullet$} (D);
	\end{tikzpicture}
	}\label{fig:network}
	}
}
\qquad
\subfloat[][\footnotesize Two possible allocations, on a channel between $A$ and $B$ that is  controlled by an attacker $E$, for the guardian $G$ when he defends an honest agent $A$. For both cases (the above one is implicit), we assume the presence of an authentic and resilient communication channel between $G$ and $A$ (dashed line).]
{
\parbox{0.45\textwidth}
	{\centering
	 \scalebox{0.8}{
	 \begin{tikzpicture}[font=\small]
	 	\node (A) {$A$};
	 	\node (B) [right=of A,xshift=75] {$B$};
	 	\draw[double,double distance=1.5pt,line width=0.5pt] (A.0) -- (B);
	 	\node[] (G) [right=of A,label=above:$G$] {$\blacksquare$};%{$\qed$};
	 	\node (E) [right=of G,label=above:$E$] {$\bullet$};
	
	 	\node (A1) [below=of A]{$A$};
	 	\node (B1) [right=of A1,xshift=75] {$B$};
	 	\draw[double,double distance=1.5pt,line width=0.5pt] (A1.0) -- (B1);
	 	\node[] (E1) [right=of A1,label=above:$E$] {$\bullet$};
	 	\node (G1) [right=of E1,label=above:$G$] {$\blacksquare$};% {$\qed$};
	
	 	\draw[->,dashed] (G1)  to [bend left=45] (A1);
	 \end{tikzpicture}
	 }\label{fig:secChannel}
	}

}
\caption{Model of the network and possible allocations of the guardian on a channel.}\label{fig:division}
\end{figure*}

We model the network as a graph (an example is depicted in 
Fig.~\ref{fig:network}), where vertices 
represent the agents of the network and edges represent communication channels 
(we assume no properties of these channels, which are standard insecure 
channels over which messages are sent as specified by the security protocols). 
Since, as we remarked above, it would be unfeasible for the guardian to
defend the traffic on all network channels, we investigate which of these
channels the guardian should be best positioned on.

Security protocols typically involve two honest agents $A$ and $B$, who
sometimes enroll also a honest and trusted third party $S$ (we could, of
course, consider protocols with more agents). As depicted in
Fig.~\ref{fig:network}, the DY-attacker $E$ is in control of all the
communication channels of the network, thus, in the case of a ping-pong
protocol between $A$ and $B$, $E$ controls also the communication
channel between $A$ and $B$. If we were to allocate a guardian $G$ on
such a channel in order to defend the honest agent $A$, it could only be
in one of two locations: as shown in Fig.~\ref{fig:secChannel}, either
the guardian $G$ is between the initiator $A$ and the attacker $E$, or
$G$ is between the attacker $E$ and the responder $B$. In the following,
these two cases will be used as a base of network topologies to be
considered during the analysis. We will see in the next section that the
guardian should have the possibility of alerting $A$ of the ongoing
attack without being detected by the attacker; in such a case (especially as 
highlighted in the lower topology in Fig.~\ref{fig:secChannel}),
we thus assume the presence of
an authentic and resilient communication channel (confidentiality can be
enforced but it is not mandatory) between $G$ and $A$.\footnote{This
channel could be a digital or a physical channel, say a text message
sent to a mobile (as in some two-factor authentication or e-banking
systems), a phone call (as in burglar alarm systems), or even a flag
raised (as is done on some beaches to signal the presence of sharks). We
do not investigate the features of this channel further but simply
assume, as done in all the above three examples of runtime guarding
(monitoring) systems, that such a channel actually exists.} In the
following, this channel will be omitted from the notation and the
figures for the sake of readability.

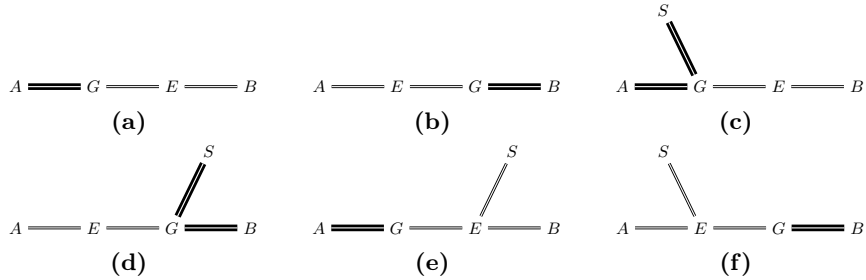
\begin{figure}[t]
\centering
	\subfloat[][\label{fig:casesA}] 
	{
	\scalebox{0.7}{
	\begin{tikzpicture}
		\node (A) {$A$};
		\node (G) [right=of A]{$G$};
		\node (E) [right=of G]{$E$};	
		\node (B) [right=of E]{$B$};
		\draw[double,line width=1.5pt] (A) -- (G);
		\draw[double,line width=0.5pt] (G) -- (E);
		\draw[double,line width=0.5pt] (E) -- (B);
		\begin{pgfonlayer}{background}
	%    \node [draw=black,densely dashed,fill=gray!10,fit=(A) (G), rounded corners] {};
		\end{pgfonlayer}
	\end{tikzpicture}
	}%scalebox
	}
	\quad
	\subfloat[][\label{fig:casesB}] 
	{
	\scalebox{0.7}{
	\begin{tikzpicture}
		\node (A) {$A$};
		\node (E) [right=of A]{$E$};
		\node (G) [right=of E]{$G$};	
		\node (B) [right=of G]{$B$};
		\draw[double,line width=0.5pt] (A) -- (E);
		\draw[double,line width=0.5pt] (E) -- (G);
		\draw[double,line width=1.5pt] (G) -- (B);
		\begin{pgfonlayer}{background}
	%    \node [draw=black,densely dashed,fill=gray!10,fit=(B) (G), rounded corners] {};
	\end{pgfonlayer}
	\end{tikzpicture}
	}%scalebox
	}
	\quad
	\subfloat[][\label{fig:casesC}] 
	{
	\scalebox{0.7}{
	\begin{tikzpicture}
		\node (A) {$A$};
		\node (G) [right=of A]{$G$};
		\node (S) [above=of G,xshift=-20pt] {$S$};
		\node (E) [right=of G]{$E$};	
		\node (B) [right=of E]{$B$};
		\draw[double,line width=1.5pt] (A) -- (G);
		\draw[double,line width=1.5pt] (G) -- (S);
		\draw[double,line width=0.5pt] (G) -- (E);
		\draw[double,line width=0.5pt] (E) -- (B);
		\begin{pgfonlayer}{background}
	%    \node [draw=black,densely dashed,fill=gray!10,fit=(A) (S) (G), rounded corners] {};
		\end{pgfonlayer}
	\end{tikzpicture}
	}%scalebox
	}
	\quad
	\subfloat[][\label{fig:casesD}]
	{
	\scalebox{0.7}{
	\begin{tikzpicture}
		\node (A) {$A$};
		\node (E) [right=of A]{$E$};	
		\node (G) [right=of E]{$G$};
		\node (S) [above=of G,xshift=20pt] {$S$};
		\node (B) [right=of G]{$B$};
		\draw[double,line width=0.5pt] (A) -- (E);
		\draw[double,line width=0.5pt] (E) -- (G);
		\draw[double,line width=1.5pt] (G) -- (S);
		\draw[double,line width=1.5pt] (G) -- (B);
		\begin{pgfonlayer}{background}
	%		\node [draw=black,densely dashed,fill=gray!10,fit=(B) (S) (G), rounded corners] {};
		\end{pgfonlayer}
	\end{tikzpicture}
	}%scalebox				
	}
	\quad
	\subfloat[][\label{fig:casesE}] 
	{
	\scalebox{0.7}{
	\begin{tikzpicture}
		\node (A) {$A$};
		\node (G) [right=of A]{$G$};	
		\node (E) [right=of E]{$E$};
		\node (S) [above=of E,xshift=20pt] {$S$};
		\node (B) [right=of E]{$B$};
		\draw[double,line width=1.5pt] (A) -- (G);
		\draw[double,line width=0.5pt] (G) -- (E);
		\draw[double,line width=0.5pt] (E) -- (S);
		\draw[double,line width=0.5pt] (E) -- (B);
		\begin{pgfonlayer}{background}
	%	\node [draw=black,densely dashed,fill=gray!10,fit=(A) (G), rounded corners] {};
		\end{pgfonlayer}
	\end{tikzpicture}
	}%scalebox
	}
	\quad
	\subfloat[][\label{fig:casesF}] 
	{
	\scalebox{0.7}{
	\begin{tikzpicture}
		\node (A) {$A$};
		\node (E) [right=of A]{$E$};
		\node (S) [above=of E,xshift=-20pt] {$S$};
		\node (G) [right=of E]{$G$};	
		\node (B) [right=of G]{$B$};
		\draw[double,line width=0.5pt] (A) -- (E);
		\draw[double,line width=0.5pt] (E) -- (S);
		\draw[double,line width=0.5pt] (E) -- (G);
		\draw[double,line width=1.5pt] (G) -- (B);
		\begin{pgfonlayer}{background}
	%	    \node [draw=black,densely dashed,fill=gray!10,fit=(B) (G), rounded corners] {};
		\end{pgfonlayer}
	\end{tikzpicture}
	}%scalebox
	}\caption[]{Base cases of network topologies for protocols between two agents (a, b) and three agents (c, d, e, f). We denote with double stretched lines (in boldface) the channels for which we assume that the attacker is not present.}
\label{img:BaseCasesOfNetworkTopologies}
\end{figure}

If the network topologies for two-agent protocols are simple
(Fig.~\ref{fig:casesA} and \ref{fig:casesB}), for the case where a
trusted third party $S$ (or another agent) is present on the network, we
have to make some assumptions about the position of the attacker $E$
(the attack power of the attacker is never questioned). In this paper,
we consider four main base cases of network topologies for three-agent
protocols, where, for every case, we consider which channel(s) the guardian
is defending: 
\begin{itemize}
\item Fig.~\ref{fig:casesC}: the channel between $A$ and $S$ (we assume
that the attacker is not present over these channels\footnote{We do not
make assumptions on the real topology of the network between $A$ and $S$
(i.e., there could be more than one channel) but only consider the fact
that the communications from $E$ are received by $G$.} and the guardian
acts like a proxy),
\item Fig.~\ref{fig:casesD}: the channel between $B$ and $S$ (this is
the specular scenario with respect to the previous case),
\item Fig.~\ref{fig:casesE}: $A$'s communication channel (the guardian
acts as a proxy for $A$), and
\item Fig.~\ref{fig:casesF}: $B$'s communication channel (the guardian
acts as a proxy for $B$).
\end{itemize}
These basic topologies abstract the communication channels of a
complex network (e.g., a LAN) in a way that permits one to reason
about the position of agents without introducing additional
parameters in the process (e.g., additional agents that start the
protocol at the same time, or multiple network paths relaxed in one
link).

In general, we cannot state that a base case is the right one or the
wrong one as this actually depends on both the analyzed protocol and the
agent we want to defend.
In order to implement the right guardian, we should consider the protocol 
defense possible in each of these cases. We conjecture that all other network 
topologies with two or three agents can be reduced to the base cases 
introduced above, but leave a formal proof for future work.

\subsection{Network guardian in practice}
Attacks leverage protocol-dependent features, and thus attack traces
always contain particular messages that we can use as signals for
ongoing attacks. 
As messages transit continuously through the network, we assume that the 
guardian has a way to distinguish them (otherwise, we cannot guarantee any 
type of defense).
In order to operate, the network guardian needs to interact with the
messages transiting over the network. The two modules that we define
in the architecture of the guardian are: (i) the \emph{Identification
Module}, and (ii) the \emph{Control Module}. Both modules operate
separately, do not interact with each other (even though they share the
guardian's dataset $D_G$), and are meant to (i) distinguish the
messages that belong to the protocol\footnote{We deliberately wrote
``protocol'' instead of ``protocols'' since, for now, we are not going
to consider multi-protocol attacks or protocol composition, e.g.,
\cite{CiobacaCortier10,CD-fmsd08,ModVig2009}. As future work, we
envision a distinguisher able to to distinguish between messages
belonging to different protocols and thus consider also the attacks that
occur when messages from one protocol may be confused with messages from
another protocol.} that they are defending and (ii) detect ongoing
attacks.

These features are achieved by means of two \emph{distinguishers}
$\DistinguisherR$ and $\DistinguisherC$, two probabilistic polynomial
time algorithms.
$\DistinguisherR$ returns $1$ if it believes that a message $m$ belongs to the protocol and $0$ otherwise.
We use the distinguisher $\DistinguisherC$ in order to detect, from the run of 
a security protocol $\mathcal{P}$ (identified by the other module), those 
messages $m$ that are considered \emph{critical}, i.e., that can be used to attack $\mathcal{P}$.

For a concrete example of critical message, we can refer to
Table~\ref{tab:isoSC27}. The nonce $N_A$ exchanged in message $(1.1)$ is
the first information that the attacker uses in order to perform the
reply attack against the ISO-SC 27 protocol, so this message must be
considered critical. Even though the nonce is sent as a plaintext, the
use of the distinguisher $\DistinguisherC$ overcomes the problem with
encrypted messages.

\begin{figure}[t]
\centering	
\subfloat[][Identification Module: Assuming that $m$ is the message spied from the spy filter, $\dState$ is the state where the distinguisher is invoked on input $m$, $\fState$ the ``forward state'', $\rState$ the ``label state''
in which the message $m$ is labeled in the dataset $D_G$ as part of the protocol $\mathcal{P}$. \label{fig:recordModule}] 
{
  \scalebox{0.75}{
	\qquad%\input{img/RecordModule}\qquad
	\begin{tikzpicture}[auto, node distance=50pt,every transition/.style={->,>=stealth'}]
	%\node[draw,rounded corners,dotted,fill=gray!60] (S) {\emph{\small{Spy-Filter}}};
		\node[draw,rounded corners,minimum width=20pt,minimum height=20pt,fill=gray!10] (D) [] {$\dState$};
		\node[draw,rounded corners,minimum width=20pt,minimum height=20pt,fill=gray!10] (F) [above=of D,xshift=40pt,yshift=-20pt]{$\fState$};
		\node[draw,rounded corners,minimum width=20pt,minimum height=20pt,fill=gray!10] (R) [right=of D,xshift=50pt]{$\rState$};
		
		%\draw[->, dotted] (S) to node [] {$m$} (D);	
		\draw[transition,->,bend right] (D) to node [swap,bend right=45] {$0$} (F);
		\draw[transition,->] (D) to node [] {$1$} (R);
		\draw[transition,->,bend right] (R) to node [swap,bend right=45] {$0/1$} (F);
		\draw[transition,->,bend right] (F) to node [swap, bend left=45] {$0/1$} (D);
		
		\begin{pgfonlayer}{background}
		    \node [draw=black,dotted,fill=gray!10,fit=(D) (R) (F), rounded corners] {};
		\end{pgfonlayer}	
	\end{tikzpicture}\qquad
  }%scalebox
}
\qquad\qquad
\subfloat[][Control Module: Assuming that $m$ is the message spied from the spy filter, $\dState$ is the state where the distinguisher is invoked on input $m$, $\iState$ the state where the attack invariant is invoked on $m$, $\fState$ represents the ``forward state'', $\infState$ the ``interference state''. %i.e. an algorithm, executed if %and only if $\\Invariant(m_i)=1$, in order to carry out a interference to neutralize an ongoing attack. 
\label{fig:controlModule}] 
{
  \scalebox{0.75}{
	\qquad%\input{img/ControlModule}
	\begin{tikzpicture}[auto, node distance=50pt,every transition/.style={->,>=stealth'}]
		%\node[draw,rounded corners,dotted,fill=gray!60] (S) {\emph{\small{Spy-Filter}}};
		\node[draw,rounded corners,minimum width=20pt,minimum height=20pt,fill=gray!10] (D) [xshift=30pt] {$\dState$};
		\node[draw,rounded corners,minimum width=20pt,minimum height=20pt,fill=gray!10,xshift=50pt] (I) [right=of D] {$\iState$};
		\node[draw,rounded corners,minimum width=20pt,minimum height=20pt,fill=gray!10] (F) [above=of D,xshift=40pt,yshift=-20pt]{$\fState$};
		\node[draw,rounded corners,minimum width=20pt,minimum height=20pt,fill=gray!10] (R) [below=of I,xshift=-40pt,yshift=20pt]{$\infState$};
		
		%\draw[->, dotted] (S) to node [] {$m$} (D);
		\draw[transition,->,bend right] (D) to node [swap,bend right=45] {$0$} (F);
		\draw[transition,->] (D)  to node [] {$1$} (I);
		\draw[transition,->,bend right] (I) to node [swap, bend right=45] {$0$} (F);
		\draw[transition,->,bend left] (I) to node [bend left=45] {$1$} (R);
		
		\draw[transition,->,bend right] (F) to node [swap, bend left=45] {$0/1$} (D);
		\draw[transition,->,bend left] (R) to node [bend left=45] {$0/1$} (D);
		
		\begin{pgfonlayer}{background}
		    \node [draw=black,dotted,fill=gray!10,fit=(D) (F) (R) (I), rounded corners] {};
		\end{pgfonlayer}	
	\end{tikzpicture}	
	\qquad
  }%scalebox
}
\caption[]{Identification and Control Modules implemented in the guardian.}
\label{fig:RecConModule}
\end{figure}
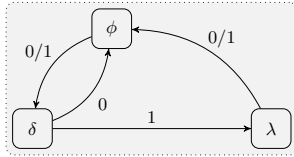
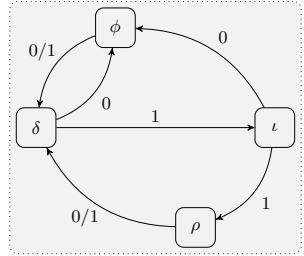

\subsubsection{Identification Module} 
Fig.~\ref{fig:recordModule} shows the graphical representation of the
\emph{Identification Module}. The guardian uses this module, together with the
distinguisher $\DistinguisherR$, to detect those messages $m$ that
belong to the protocol and label them as part of $\mathcal{P}$ in the dataset $D_G$ in order to do inference subsequently.

We can see the Identification Module as a finite state machine where the transition from state to state depends on the spied messages.
When a message $m$ is spied by the spy filter (see Table~\ref{OurDYmodel} for 
the two available spy filters), the Identification Module of the guardian 
invokes the distinguisher $\DistinguisherR(m)$ to establish whether the message belongs to the protocol or not.

If $\DistinguisherR(m) = 0$, the message is not considered useful
and the guardian moves to the forward state $\fState$, which will
let the message go, and subsequently goes back, without checking
any condition, to the initial state $\dState$ in order to wait for
the next message. If $\DistinguisherR(m) = 1$, then $m$ belongs to
the protocol and the guardian moves to the ``identification state''
$\rState$, where the message is labeled in the dataset $D_G$. After
the message has been labeled, the Identification Module goes back
to the initial state $\dState$ in order to wait for the next
message.

From now on, when we do an operation (spy-filters excluded) on the
dataset, we mean (slightly abusing notation) the subset of the
labeled messages.

\subsubsection{Control Module}
Fig.~\ref{fig:controlModule} shows the graphical
representation of the Control Module. The guardian uses this
module, together with the distinguisher $\DistinguisherC$, in order
to deal with those messages $m$ that he must control in order to be
able to do inference (i.e., check if an attack is ongoing) and
eventually interfere with the attacker; we call these messages
\emph{critical}.

% Again, we use a distinguisher implemented in the Control Module.
Once the distinguisher, implemented in the Control Module,
believes that $m$ is critical (at time $i$), the \emph{attack invariant} $\Invariant (m, i)$
is tested to discover (or exclude) an ongoing attack. $\Invariant (m, i)$ is a 
protocol-dependent Boolean condition; formally, it is a first-order logic 
formula on a critical message of the protocol (which can be straightforwardly 
extended to a set of messages) tested at time $i$ (i.e., after $i$ actions on the dataset $D_{net}$; in order to define more complex functions, more than two  parameters can be used):
\begin{displaymath}
	\Invariant(m, i) =
		\begin{cases}
			1 & \text{if $m$, at time $i$, characterizes an ongoing attack or a false} \\
			& \text{positive} \\
			0 & \text{if $m$, at time $i$, characterizes a normal run or a false negative} 
	\end{cases}
\end{displaymath}
If the computation of the invariant returns $1$, then the guardian $G$
carries out the appropriate defense for the attack making the victim
abort the current run of the protocol and, eventually, mislead the
attacker and/or induce him to abort the attack. We give an example of
invariant in Section~\ref{ISOsc27} when we return to our case study.

When a message $m$ is spied by the spy filter, the Control Module is in
the initial state $\dState$, and then the message is passed as input to
the distinguisher $\DistinguisherC$, whose task is to establish whether
the message is critical or not. If the result of the distinguisher is
$\DistinguisherC(m) =0$, the message is not considered critical and the
guardian moves to the forward state $\fState$, which will let the
message go, and subsequently goes back, without checking any condition,
to the initial state $\dState$ in order to wait for the next message.
Instead, if $\DistinguisherC(m) =1$, then a critical message has just
been distinguished from the others; the guardian moves to the invariant
state $\iState$ passing the message as input to the attack invariant
formula $\Invariant(m, i)$, whose task is to establish whether an attack is
actually ongoing or not (the invariant is computed using the labeled
messages in $D_G$ respecting the temporal constraints). If
$\Invariant(m, i) = 0$, then either an attack is not ongoing or a false
negative has just happened (i.e., the defense mechanism is partial);
thus, the guardian goes to the forward state $\fState$, which will let
the message go, and subsequently goes back without checking any
condition to the initial state $\dState$. Instead, if $\Invariant(m, i) =1$
either an attack is ongoing or a false positive has just happened,
independently of the used defense mechanism; thus, the guardian moves
to interference state $\infState$ to carry out the appropriate
countermeasures and subsequently goes back, without checking any
condition, to the initial state.

As the $\DistinguisherR$ is needed in order to detect the messages that
belong to the protocol $\mathcal{P}$, we envision $\DistinguisherC$ to
be useful in the case of protocols with a large number of messages
% exchanged by agents, 
in order to lighten the computation load of $\Invariant(m,i)$, i.e., we
compute $\Invariant(m, i)$ on a subset of the protocol messages:
\[
\mathit{Critical} \quad \subseteq \quad \mathcal{P}_\mathit{labeled} \quad \subseteq \quad \mathit{Messages}
\]
where $\mathit{Messages}$ are all the messages saved in the dataset by
a spy-filter, $\mathcal{P}_\mathit{labeled}$ are the messages that
$\DistinguisherR$ labeled as part of the protocol $\mathcal{P}$ and
$\mathit{Critical}$ are the messages that $\DistinguisherC$ believes
may be used to attack $\mathcal{P}$. 
% Since $\Invariant(m)$ in computed
% with $m\in \mathit{Critical}$, protocol composition and multi-protocol
% attacks are not in the scope of this paper, as we remarked before.

\subsection{Topological advantage}

% To defend protocols against attacks, a guardian should be near the
% victim(s). If a guardian is not ``near'' the agent that he is defending,
% then he could be useless: if he does not see (and thus cannot control)
% messages in transit from two (or more) agents, then he cannot carry
% out the defense/interference.

To defend protocols against attacks, a guardian should be ``near''
one of the agents involved in the protocol executions; otherwise
the guardian could be useless: if he does not see (and thus cannot
control) messages belonging to the protocol in transit from these
agents, then he cannot carry out the interference/defense.

% \fix{M}{paragrafo riscritto per essere più aderenti alla
% definizione}

\begin{definition}[Topological Advantage]\label{def:top-advantage}
Let $X\in\Agents{}$ be the agent that the guardian $G\in\bEves$ is
defending in a particular protocol (with set $\mathit{Messages}$ of messages), and $Y\in\Agents{}$ the other agent.
We say that $G$ is in \emph{topological
advantage} with respect to the attacker $E$ if
\begin{align*}
\forall m\in & \mathit{Messages}.\ \exists i\in \mathbb{N}. \,\\
& G \in \canseeM{<\!X, m, Y\!>, i} \ \lor \ G \in \canseeM{<\!Y, m, X\!>, i} \ \lor\\
& G \in \canseeM{<\!E(X), m, Y\!>, i} \ \lor \ G \in \canseeM{<\!Y, m, E(X)\!>, i}
\end{align*}
\end{definition}

Definition~\ref{def:top-advantage} states that for a guardian to be
in topological advantage, he must be collocated over the network in
one of the configurations of
Fig.~\ref{img:BaseCasesOfNetworkTopologies} so that he can spy (and
eventually modify) all the transiting messages to and/or from the agent that 
he is defending, even in the case that they are forged.

In order to define what a defense mechanism is, we have to
formalize how an attack can be formalized based on a parametric
function that the attacker computes during his execution.

Let $E \in \Eves$, $X \in \Honest$, $s$ be the number of steps
composing the attack trace, $m_s$ the message spied over the
network or present in the attacker dataset $D_{E}$ at step $s$,
$\mathit{Func}=\{\mathit{Erase}, \mathit{Injection},
\mathit{Duplicate}, \ldots\}$ a set of functionalities that $E$ can
use on the messages. Note that the names of the functionalities
quite intuitively denote their meaning; not all of the
functionalities are used in this paper and many more could be
defined. 
The functionalities in $\mathit{Func}$ have domain in
the messages belonging to a given protocol, whereas the codomain is defined as the union of all the possible transformations 
of  the messages in the domain that give (i)  messages ``acceptable'' by the protocol (i.e., that can be sent/received by the protocol's agents) or (ii) an empty message. 
The codomain is thus a set of messages. 
We use ${\mathit{func}}_s$ to denote a functionality in $\mathit{Func}$ 
used at step $s$. 

\begin{definition}[Attack Function]\label{def:att-function} 
The attack function $f(m, s)$ selects a functionality $\mathit{func}_s$ 
to be used on the message $m$ at step $s$ and returns the result of the   $\mathit{func}_s$ with argument $m$ ($\mathit{func}_s(m)$).
\end{definition}
As a concrete example, the attack function of the attack in Table~\ref{tab:isoSC27} is:
\begin{center}
\begin{tabular}{C{0.6cm}C{2.5cm}C{2.5cm}C{2.5cm}}
 \hline
 $s$ & $m$ & $\mathit{func}_s$ & $f(s, m)$\\%& Agent\\
 \hline
 $1$ & $N_A$ & $\mathit{Erase}$ & $\emptyset$ \puntello \\
 $2$ & $N_A$ & $\mathit{Injection}$  & $N_A$ \puntello \\
 $3$ & ${\{\!|N_A,N'_A|\!\}}_{K_{AB}}$ & $\mathit{Erase}$ & $\emptyset$  \puntello\\
 $4$ & ${\{\!|N_A,N'_A|\!\}}_{K_{AB}}$ & $\mathit{Injection}$ & ${\{\!|N_A,N'_A|\!\}}_{K_{AB}}$ \puntello\\
 $5$ & $N'_A$ & $\mathit{Erase}$ & $\emptyset$ \puntello\\
 $6$ & $N'_A$ & $\mathit{Injection}$ & $N'_A$ \puntello \\
\hline
\end{tabular} 
\label{attackFunctionTable}
\end{center} 

Of course, more complex attack functions could (and sometimes even
should) be defined, especially for more complex protocols. Since
the attack function is but one parameter, we believe that our
definitions and results are general enough and can be quite easily
adapted to such more complex functions.

Having formalized how an attack can be seen as a parametric
function, we can also assume the existence of an inverse function
$f^{-1}(m, s)$ of the attack function (i.e., the function that from
a message $m$ such that $m=f(m', s)$, and a step $s$, computes $m'$). In this
paper, we will not discuss how to formalize the inverse attack
function; we leave a definition for future work and for now assume
that, during the implementation of the framework, a security
analyst can take care of this matter.

\begin{definition}[Defense Mechanism]\label{def:def-mechanism}
Let $X\in\Agents{}$ be the agent that the guardian $G\in\bEves$ is
defending in a particular protocol (with set $\mathit{Critical}$ of
critical messages), let $E\in\Eves$ be the attacker, and $s$ be the number 
of steps composing $E$'s attack trace. We say that $G$ is a \emph{defense mechanism} 
if he knows $E$'s attack function $f(m, s)$ and can compute the inverse function $f^{-1}(m,s)$ in order to enforce the following:
\begin{multline*}
\nexists m \in \mathit{Critical}.\, 
\forall i\in\mathbb{N}.\  \exists p, j\in\mathbb{N}.\ j > i \ \wedge \ 1\le p \le s \ \wedge \\
m\in D^{i}_{net} \ \wedge
f^{-1}(f(m, p), p)=m \ \wedge \\
(G\notin \canseeM{<\!E, f(m, s), X\!>, j} \lor G\notin \canseeM{<\!E(Y), f(m, s), X\!>, j})
\end{multline*}
\end{definition}

If $G$ can compute the inverse attack function, then $G$ has
knowledge of the possible attacks against the protocol carried out through the attack function and can detect the critical messages even if the attacker modifies/deletes them.

Thus, we can state the following theorem (which can be quite straightforwardly generalized to multiple attackers):
\begin{theorem}\label{thm:defense}
A guardian $G\in\bEves$ is a defense mechanism for an agent $X\in\Agents{}$ 
in a protocol $\mathcal{P}$, %with an agent $Y\in\Agents{}$ 
if he is in topological advantage with respect to an attacker $E\in\Eves$ who is attacking $X$ in $\mathcal{P}$.
\end{theorem}

As a proof sketch, let $X\in\Agents{}$ be the agent that $G$ is defending,
$Y\in\Agents{}$, $E\in\Eves$ with attack function $f(m, p)$, $m\in \mathit{Critical}$, $f^{-1}$ known to $G$, $G$ in topological advantage with respect to the attacker $E$, $s$ the number of steps composing $E$'s attack trace, and $1 \le p\le s$. Then, since $f(m, p)\in  \mathit{Messages}$, we have that:
$\exists i\in \mathbb{N}. \
G \in \canseeM{<\!X, f(m, p), Y\!>, i} \ \lor \ G \in \canseeM{<\!Y, f(m, p), X\!>, i} \ \lor \
G \in \canseeM{<\!E(X), f(m, p), Y\!>, i} \ \lor \ G \in \canseeM{<\!Y, f(m, p), E(X)\!>, i}
$.
In order to have a defense mechanism, we have to enforce the following:
$\nexists m \in \mathit{Critical}.\
\forall i\in\mathbb{N}.\  \exists p, j\in\mathbb{N}.\ j > i \ \wedge \ 1\le p \le s \ \wedge \
m\in D^{i}_{net} \ \wedge \
 G\notin \canseeM{<\!E, f(m, p), X\!>, j} \wedge
f^{-1}(f(m, p), p)=m
$.
Since $f(m, p)\in \mathit{Critical} \subseteq \mathit{Messages}$,  only $f^{-1}(f(m, p),p)=m$ must be enforced, but it is known to $G$ by assumption.

\section{Case studies}
\label{sec:case-studies}

\subsection{The ISO-SC 27 protocol}\label{ISOsc27}

Even though the ISO-SC 27 protocol is subject to the parallel sessions
attack shown in Table~\ref{tab:isoSC27}, we can defend it by means of
a guardian $G$.
Since the victim is $A$, for the defense to be possible, it is necessary
that $G$ is in the configuration in Fig.~\ref{fig:casesA},
i.e., between $A$ and the rest of the network agents, so that he can
identify/control all of $A$'s incoming and outgoing messages (by 
Definition~\ref{def:top-advantage}, in this configuration the guardian is in topological advantage), whereas in the configuration in Fig.~\ref{fig:casesB} 
he can be completely excluded by an attacker $E$. 
In the following, we give as an example the successful case and a brief 
explanation for the unsuccessful one. 

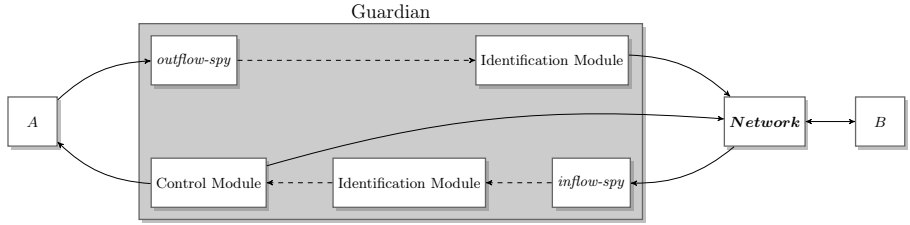
\begin{figure*}[t]
\centering
\scalebox{0.62}{
	\begin{tikzpicture}[node distance=80pt, every transition/.style={->,>=stealth'}]

	\node[drop shadow,thick,draw={black!60},fill=white, minimum width=30pt, minimum height = 30pt] (A) {$A$};
	\node[drop shadow,thick,draw={black!60},fill=white, minimum width=30pt, minimum height = 30pt] (Outflow) [above right=of A,yshift=-50pt]{\emph{outflow-spy}};
	\node[drop shadow,thick,draw={black!60},fill=white, minimum width=30pt, minimum height = 30pt] (Id1) [right=of Outflow,xshift=65pt]{Identification Module};
	\node[drop shadow,thick,draw={black!60},fill=white, minimum width=30pt, minimum height = 30pt] (Control) [below right=of A,yshift=50pt]{Control Module};	
	\node[drop shadow,thick,draw={black!60},fill=white, minimum width=30pt, minimum height = 30pt] (Id2) [right=of Control,xshift=-40pt]{Identification Module};
	\node[drop shadow,thick,draw={black!60},fill=white, minimum width=30pt, minimum height = 30pt] (Inflow) [right=of Id2,xshift=-40pt]{\emph{inflow-spy}};

	\node[drop shadow,thick,draw={black!60},fill=white, minimum width=30pt, minimum height = 30pt] (NETWORK) [above right=of Inflow,yshift=-50pt]{\bm{$Network$}};	
	\node[drop shadow,thick,draw={black!60},fill=white, minimum width=30pt, minimum height = 30pt] (B) [right=of NETWORK,xshift=-50pt]{$B$};

	\draw[transition] (A) to [bend right=-20] (Outflow);
	\draw[transition] (Id1) to [bend right=-20] (NETWORK);
	\draw[transition,dashed] (Outflow) to [] (Id1);
	\draw[transition,dashed] (Inflow) to [] (Id2);
	\draw[transition,dashed] (Id2) to [] (Control);
	\draw[transition] (Control) to [bend right=-20] (A);
	\draw[transition] (Control) to [bend right=-10] (NETWORK);
	\draw[transition] (NETWORK) to [bend right=-20] (Inflow);

	\draw[transition,<->] (NETWORK) -- (B);

	\begin{pgfonlayer}{background}
		\node[drop shadow,thick,draw={black!60},fill=white, minimum width=30pt, minimum height = 30pt][fill=gray!40, inner sep=7pt,fit=(Outflow) (Inflow), label=\large{Guardian}] {};
	\end{pgfonlayer}		
	\end{tikzpicture}
}
\caption[]{Guardian configuration for the ISO-SC 27 protocol. With a dashed arrow we describe  the fact that the execution flow (not the spied message) continues with the next module. \label{fig:gConf}}
\end{figure*}

In  order to defend the ISO-SC 27 protocol, we have set up the guardian $G$ 
with the two spy-filters shown in Fig.~\ref{fig:gConf}: an 
outflow-spy filter in order to record in his dataset $D_G$ all of $A$'s 
outgoing messages, and an inflow-spy filter in order to record and control $A$'s incoming messages.

Even if $G$ does not know the symmetric key $K_{AB}$, he can become
aware that the protocol has been attacked when he spies via the
\emph{inflow-spy} filter a message of the same form of the message (1) in 
Table~\ref{tab:isoSC27} (i.e., $N_A$; the guardian knows that the attacker 
will reply the first message because he knows the attack function of 
Definition~\ref{def:att-function}) between those that have
previously been identified as such: 
if an attack is ongoing, then the message that has been identified by the 
Control Module as critical (i.e., is one of the first messages of the protocol) ``has already been seen'' by $G$. 
We formalize this concept by means of the invariant $\Invariant(m, i)$:
\begin{displaymath}
 \exists{m'} \in D^{i-1}_G.\ \DistinguisherC{(m)}=1 ~\wedge~  
 \DistinguisherC{(m')}=1 \ ~\wedge~ m = m' .
\end{displaymath}
That is, if an attack is ongoing and $m$ is the message spied by guardian's inflow-spy filter, labeled by the Identification Module, and in the Control Module the distinguisher $\DistinguisherC$ believes that it is critical, then the guardian's dataset $D_G^{i}$ must
contain another message $m'$ seen before such that $m=m'$ (the implementation of $D_G$ must be done with respect to the temporal constraints of the invariant $\Invariant{}$, but in this paper we do not discuss the implementation details).
Since the guardian knows that the attacker can use a replay attack, by 
Definition~\ref{def:def-mechanism}, he has to define the inverse of the attack 
function as the identity function (the use of the identity function is also reflected in the definition of the invariant).\footnote{Formally, for the ISO-SC 27 we have: $f^{-1}(f(N_A,2),2) = f^{-1}(N_A,2) = N_A$ (where $s = 2$ refers to message $(2)$ in Table~\ref{attackFunctionTable} or, equivalently, message $(1.2)$ in Table~\ref{tab:isoSC27}).}

Let us assume, following~\cite{FPV:02SECRYPTbook,FPV:SECOTS}, that each
honest agent defended by the guardian $G$ has a set of flags that $G$
can modify in order to make the agent he is defending abort the protocol. Once 
he has detected such an ongoing attack, $G$ can defend it carrying out the 
interference. He modifies the content (i.e., he alters the nonce $N_A$) of the 
first message in the parallel session (see Table~\ref{isoRET} for the complete 
execution trace, and Table~\ref{tbl:isosc27} for the corresponding dataset evolution). 
At this point, the guardian already knows that an attack is ongoing, but we 
choose to finish the two sessions of the protocol ($G$ changes $A$'s ``abort 
flag'' only at the end) in order to show that we can also deliver false 
information to the attacker and that the Control Module (shown in 
Table~\ref{tbl:isosc27}) checks the invariant only once since the replayed 
message in (1.2) is not seen as critical (i.e., it has not the form of the 
first message). More specifically, Table~\ref{isoRET} shows the interference 
attack that $G$ can use against 
the attacker $E$, and Table~\ref{tbl:isosc27} the evolution of the dataset 
and the inference during the protocol execution.

\begin{table}[t!]
	\caption{Guardian's interference for the ISO-SC 27 protocol. \label{isoRET}}
\begin{center}
\scalebox{1.0}{
\begin{tabular}{c}
 \toprule
	{\bf Interference}\\
 \midrule
 $
\begin{array}{rll}
	(1.1\phantom{_1}) & A    \to E(B)    & : N_A \\
	(2.1\phantom{_1}) & E(B) \to G(A)    & : N_A \\
	(2.1_1) 	 	  & G(B) \to A       & : N_\mathit{fake} \\	
	(2.2\phantom{_1}) & A    \to E(B)    & : \{\!| N_\mathit{fake}, N'_A |\!\}_{K_{AB}}\\
	(1.2\phantom{_1}) & E(B) \to A 	     & : \{\!| N_\mathit{fake}, N'_A |\!\}_{K_{AB}}\\
	(2.2\phantom{_1}) & G \text{ raises } A\text{'s flag for abort}
\end{array}
 $
\\
\bottomrule
\end{tabular}
}
\end{center}
\end{table}

\begin{table}[t]
\caption[]{Dataset evolution and inference for the ISO-SC 27 protocol.
$\{(x.y)\}$ refers to the message sent in step $(x.y)$ (we omit the repeated messages) and to the configuration in Fig.~\ref{fig:casesA}.}
\label{tbl:isosc27}
\begin{center}
\scalebox{0.83}{
\begin{tabular}{C{.4cm}llC{2.6cm}C{1.3cm}C{1.3cm}}
 \toprule
$\bm{i}$\qquad\  & \textbf{Protocol message} & $\bm{D_G^i}$ & \textbf{Identification Module} & \multicolumn{2}{C{2.6cm}}
{\begin{tabular}[x]{@{}c@{}}\textbf{Control}\\\textbf{Module}\end{tabular}}\\%{\textbf{Control Module}}\\	
& & & $\DistinguisherR(m)$ & ${\DistinguisherC(m)}$ & ${\Invariant(m, i)}$\\
\midrule
$0$ & $-$ & 
$\{~\}$ & $-$ & $-$ & $-$\\
$1$ & $(1.1\phantom{_1}) \,\, A \to E(B) : N_A$ & 
$\{(1.1)\}$\quad & $1$ & $-$ & $-$\\
$2$ & $(2.1\phantom{_1}) \,\, E(B) \to G(A) : N_A$ & 
$\{(1.1)\}$ & $1$ & $1$ & $1$\\
$3$ & $(2.1_1) \,\, G(A) \to A : N_{fake}$ & 
$\{(1.1), (2.1_1)\}	$ & $-$ & $-$ & $-$\\
$4$ & $(2.2\phantom{_1}) \,\, A \to E(B) : \{\!|N_{fake},N_A'|\!\}_{K_{AB}}$ & 
$\{(1.1), (2.1_1), (2.2)\}$ & $1$ & $-$ & $-$\\
$5$ & $(1.2\phantom{_1}) \,\, E(B) \to A : \{\!|N_{fake},N_A'|\!\}_{K_{AB}}$ \qquad & 
$\{(1.1), (2.1_1), (2.2)\}$ & $1$ & $0$ & $-$\\
$6$ & $(2.2\phantom{_1})\,\, G \text{ raises } A\text{'s flag for abort}$ & $-$ & $-$ & $-$ & $-$\\
\bottomrule
\end{tabular}
}%scalebox
\end{center}
\end{table}

To measure the defense mechanism implemented by the guardian for the
parallel sessions attack against the ISO-SC 27 protocol, we consider false 
positives and negatives.

\paragraph{False positives:} False positives are possible if, after $A$ 
completes a protocol run as initiator, $B$ restarts the protocol with $A$ (i.e., they change roles) using (in the first message) a nonce $N_B$ that is already contained in $G$'s dataset. If $N_B$ is represented through a $k$-bit length string, then the probability of this event is equal to the probability of guessing a nonce amongst those belonging to $D^i_G$ (i.e., $G$'s dataset after $i$ actions):
\begin{displaymath}
Pr[N_B \in_R \{0,1\}^k, N_B \in D^i_G] = \frac{|D^i_G|}{2^k}
\end{displaymath}
So, this probability is negligible if $k$ is large enough (e.g., $k =
1024$).

\paragraph{False negatives:} False negatives are not possible, since not
knowing $K_{AB}$ the only way to attack the protocol with the classical
attack (Table~\ref{tab:isoSC27}) is to reflect $A$'s messages in a
parallel session; but if this situation happens, then the guardian has
already seen the message that is coming back to $A$, and thus he can
detect (and afterwards defeat) the ongoing attack. Since $G$ does not
admit false negatives for this scenario,  $G$ is a total defense
mechanism when he is in a topological advantage with respect to his
competitor(s), i.e., when he is defending $A$.

Now that we have seen the successful case, let us focus on the
configuration of Fig.~\ref{fig:casesB}. In this configuration, a
guardian would not work because $B$'s participation is not mandatory to
attack the protocol and thus $E$ can easily exclude $G$ from the run of
the protocol; thus there are no false positives and there are only false
negatives. In this case, the presence of the resilient channels does not
help because $G$ is completely excluded from seeing the execution of
the protocol and the attack.

Summing up the analysis of the case study, we have seen how a flawed
protocol as the ISO-SC 27 can be defended through the use of a guardian.
The first step of our analysis was the attack typically found via model
checking and the classical approach. We used the classical attack in
order to select the critical messages that the attacker exploits during
the attacks. Knowing the critical messages allows us to formalize the
invariant, which is also used in order to set up filters and module
configurations in the guardian architecture. Finally, we have
investigated the different outcomes with respect to the position of the
guardian in the network topology.

\subsection{Other protocols}
We have applied our approach also to a number of other security
protocols.
\begin{SPW14}
Table~\ref{tbl:protocols} summarizes our results, while a
more detailed analysis can be found in~\cite{PVZ-TR14}.\fix{Luca}{Complete the bib entry with the full details about the arxiv paper!}
\end{SPW14}
\begin{LONG}
Table~\ref{tbl:protocols} summarizes our results, while a
more detailed analysis can be found in the appendix.
\end{LONG}
For each protocol, in the table we report if the defense is total or
partial, which agent is being defended, and the topologies that
permit the defense.

In Table~\ref{tbl:protocols}, we show only the successful results
for each protocol in the given task (i.e., defending one of the
agents for the corresponding protocol). The outcome of the analysis
of these $7$ ($4$ two-agent and $3$ three-agent) protocols is quite
promising since we have a total defense in $5$ cases and a partial
defense in the remaining $2$ cases.

\begin{table}[t]
	\centering
\caption[]{Other case studies. See~\cite{boydMathuria,Clark97:asurvey} for details on the protocols.}
\label{tbl:protocols}
\begin{center}
\scalebox{1.0}{
\begin{tabular}{L{4cm}C{1.8cm}C{2cm}C{2cm}}
	\toprule                        
\textbf{Protocol} & \textbf{Defense} & \textbf{Agent Defended} & \textbf{Topology} \\
\hline 
ISO-SC 27 & Total  & $A$ & Fig.~\ref{fig:casesA}\\ 
SRA3P & Total   & $A$ & Fig.~\ref{fig:casesA}\\
Andrew Secure RPC & Partial & $A$ & Fig.~\ref{fig:casesA}\\
Otway-Rees & Total   & $A$ & Fig.~\ref{fig:casesC},~\ref{fig:casesE}\\
%& Total   & $A$~ Fig.~\ref{fig:casesE},~\ref{fig:casesC}\\\hline
Encrypted Key Exchange 
& Total   & $A$ & Fig.~\ref{fig:casesA}\\
SPLICE/AS & Total   & $A$ & Fig.~\ref{fig:casesC}\\
Modified BME & Partial & $B$ & Fig.~\ref{fig:casesD}\\
	\bottomrule
\end{tabular}
}%scalebox
\end{center}
\end{table}

\section{Conclusions and future work}
\label{sec:conclusions}

Discovering an attack to an already largely deployed security protocol
remains nowadays a difficult problem. Typically, the discovery of an
attack forces us to make a difficult decision: either we accept to use
the protocol even when knowing that every execution can potentially be
attacked and thus the security properties for which the protocol has
been designed can be compromised at any time, or we do not (generating
consequently, kind of a self denial of service). Both choices are
extreme, and typically the classical (and conservative) mindset prefers
to ``dismiss'' the protocol and hurry up with the deployment of a new
version hoping to be faster than those who are attempting to exploit the
discovered flaw.

The above results contribute to showing, we believe, that
non-collaborative attacker scenarios, through the introduction of a guardian, 
provide the basis for the active defense of flawed security
protocols rather than discarding them when the attack is found.
Regarding the concrete applicability of this approach to security
protocols, on one hand, we can use our previous work~\cite{FPV:02SECRYPTbook,FPV:SECOTS} as an approach for discovering
how two attackers interact in non-collaborative scenarios and what
type of interference the guardian can use, and, on the other hand, in
this paper we have given the means to understand how to exploit the
interference from a topological point of view, thus bringing the
guardian close to real implementation, which is the main objective of
our current work.

We are also working on a number of relevant issues, such as how the
content of, and the meaning that the honest agents assign to, critical
messages may have an influence on the defense mechanisms enforced by the
guardian, or such as how to define general attack functions and their inverses. We are also
investigating criteria that will allow us to reason about the minimal
and/or optimal configurations for protocol defenses. For instance, to
show that no further configurations are possible (by showing how $m$
possible configurations can be reduced to $n<m$ base ones, such as the
$6$ we considered here) or that the considered configuration is optimal
for the desired defense (and thus for the implementation of the
guardian). It seems obvious, for example, that Fig.~\ref{fig:casesA} is
the optimal configuration for defending the initiator $A$ in the
majority of two-agent protocols. Similarly, our intuition is that a
guardian (with an appropriate defense for a particular protocol) put in
configuration~\ref{fig:casesE} is also valid for the
configuration~\ref{fig:casesC} (and similarly for
configuration~\ref{fig:casesF} with respect to
configuration~\ref{fig:casesD}).

We envision the some general, protocol-independent results might be
possible but that ultimately both the notion (and agents' understanding)
of critical message and that of defense configuration will depend on the
details of the protocol under consideration and of the attack to be
defended against. Our hope is thus to obtain parametric results that can
then be instantiated with the fine details of each protocol and attack.

\bibliographystyle{abbrv}
\bibliography{bibliography}

\begin{LONG}
	
\appendix

\section{Other case studies}
\label{app:caseStudies}

In this appendix, we summarize  how the guardian works on the protocols reported in Table~\ref{tbl:protocols}. Sometimes, as we saw before, the control module must check the invariant $\Invariant{}$ respecting temporal constraints. In this appendix, we have two examples in which it is mandatory to respect temporal constraints: the Boyd-Maturia Example and the SPLICE/AS protocol. In these two specific cases with “respect” we mean that $G$ must check $\Invariant{}$ only on messages added to $D_G$ in the last $j$ steps of dataset evolution. This means that we are considering a “temporal window” expressed in the $\Invariant{}$ including the following constraint $\exists j . ~j < i \land m \in D_G^{i-1} \setminus D_G^{j}$. We also assume that messages are continuously added to $D_G$ (and thus the temporal window continues to move forward), otherwise this simple solution would be useless, since the last added message could be “very old” and this might allow, of course, attacks that we are trying to defeat. 

\subsection{Shamir-Rivest-Adleman Three Pass Protocol}

The Shamir-Rivest-Adleman Three Pass protocol (Table~\ref{SRA3P} and~\cite{Clark97:asurvey}) can be 
attacked by $E$ sending message (1) back to $A$ at the second step of the 
protocol in order to induce $A$ to send the last message of the protocol 
unencrypted (the protocol assumes that the cryptography employed is 
commutative). 
The guardian can defend $A$ if %and only if 
the considered scenario is the base case in Fig.~\ref{fig:casesA} setting up 
two spy filters on $A$: (i) an outflow-spy filter in order to record in his 
dataset all of $A$'s outgoing messages, and (ii) an inflow-spy filter in order to 
control whether amongst all of $A$'s incoming messages there is some message 
that is already in $D_G$. 
The interference consists in modifying the critical message in transit for $A$ 
with a random message $M_\mathit{fake}$, % of a different form
and in sending to $E(B)$ a random message $M'_\mathit{fake}$ in order to mislead him 
(this message can contain false information to be delivered to $E$). At the 
end of the protocol run, $E$ has the wrong secret. % $M$. 

For this attack, false positives are possible if some agent $B$ starts the 
protocol (at time $i$) with $A$ using a $M$ such that $\{M\}_{K_B} \in D^i_G$, 
while false negatives are not possible because to attack the protocol $E$ must 
send the message (1) back to $A$, but if this situation occurs, message (1) 
already belongs to $D_G$ and with message (2) the attack can be detected by 
$G$. A guardian in the configuration of Fig.~\ref{fig:casesB} would not work, because $B$'s participation is not mandatory to attack the protocol and thus $E$ can easily exclude $G$ from the run of the protocol.

\subsection{Andrew Secure RPC Protocol}

The Andrew Secure RPC protocol (Table~\ref{AsRCPP} and~\cite{Clark97:asurvey}) can be attacked by an 
attacker $E$ that sends the message (2) back to $A$ at step (4) of the 
protocol. The guardian can defend $A$ if %and only if 
the considered scenario is the base case in Fig.~\ref{fig:casesA}, setting 
up only an inflow-spy filter on $A$ in order to record and control all of $A$'s 
incoming messages of the form of the second or the third message.
The interference consists in modifying the critical message in transit with a 
random message $M_\mathit{fake}$ (we assume that $G$ wants to conclude the protocol 
before changing $A$'s ``abort flag''). 
At the end of the protocol run, $G$ makes $A$ abort the protocol and $E$ thinks 
of having attacked the protocol. 
For this attack, there are two scenarios that lead to false positives supposing 
that $K_{AB}$ has not been changed yet: (i) $A$ starts the protocol with an 
old nonce $N_A$ and $B$ replies with an old nonce $N_B$, or (ii) $B$ generates 
a random message (4) such that $K_{AB} = N_A + 1$ and $N_B' = N_B$. 
False negatives are also possible when $E$ attacks the protocol with an old 
message (4) that does not belong to $D_G$; however, in this case, the attack 
works only once because message (4) is recorded in $D_G$. Even though $B$'s 
participation is mandatory in the second message (we assume no-one else knows 
$K_{AB}$), a guardian in the configuration of Fig.~\ref{fig:casesB} would not 
work. He can, of course, see the second message but since the last message has 
nothing related to the previous ones, $E$ can masquerade as $B$ and attack the 
protocol replaying the second (or an old intercepted message) without $G$ can 
detect the attack.

\begin{table}[t]
\caption{Shamir-Rivest-Adleman Three Pass Protocol.\label{SRA3P}}
\begin{center}
\scalebox{1}{
\begin{tabular}{cc}
\toprule
{\bf Protocol} & {\bf Classical Attack}\\
\midrule
\\
$
\begin{array}{rll}
	(1) & A \to B & : \{\!| M |\!\}_{K_A} \\
	(2) & B \to A & : \{\!| \{\!| M |\!\}_{K_A} |\!\}_{K_B} \\
	(3) & A \to B & : \{\!| M |\!\}_{K_B} \\
\end{array}
 $
&
$
\begin{array}{rll}
	(1) & A    \to E(B) & : \{\!| M |\!\}_{K_A} \\
	(2) & E(B) \to A    & : \{\!| M |\!\}_{K_A} \\
	(3) & A    \to E(B) & : M \\
\end{array}
$
\\\\
\toprule
\multicolumn{2}{c}{
{\bf Interference}}\\
\midrule
\\
\multicolumn{2}{c}{
$
\begin{array}{rll}
	(1\phantom{_1}) & A    \to E(B) & : \{\!| M |\!\}_{K_A} \\
	(2\phantom{_1}) & E(B) \to G(A) & : \{\!| M |\!\}_{K_A} \\
	(2_1) 		    & G(A) \to A    & : M_\mathit{fake} \\
	(3\phantom{_1}) & G(A) \to E(B) & : M'_\mathit{fake} \\
	(3\phantom{_1}) & \multicolumn{2}{l}{A\text{ aborts}}
\end{array}
$
}
\\\\
\bottomrule
\end{tabular}
}
\end{center}
\end{table}

\begin{table}[t]
\caption{Andrew Secure RPC Protocol.\label{AsRCPP}}
\begin{center}
\scalebox{1.0}{
\begin{tabular}{cc}
\toprule
{\bf Protocol} & {\bf Attack trace 1} \\
\midrule 
\\
 $
\begin{array}{rll}
	(1) & A \to B & : A,\{\!|N_A|\!\}_{K_{AB}} \\
	(2) & B \to A & : \{\!|N_A+1,N_B|\!\}_{K_{AB}}\\
	(3) & A \to B & : \{\!|N_B+1|\!\}_{K_{AB}}\\
	(4) & B \to A & : \{\!|K'_{AB},N'_B|\!\}_{K_{AB}}\\
\end{array}
 $
&
$
\begin{array}{rll}
	(1) & A    \to B  	& : A,\{\!|N_A|\!\}_{K_{AB}} \\
	(2) & B    \to A    & : \{\!|N_A+1,N_B|\!\}_{K_{AB}}\\
	(3) & A    \to E(B) & : \{\!|N_B+1|\!\}_{K_{AB}}\\
	(4) & E(B) \to A    & : \{\!|N_A+1,N_B|\!\}_{K_{AB}}\\
\end{array}
$
\\\\
\toprule
{\bf Attack trace 2} & {\bf Interference 1}\\
\midrule 
\\
$
\begin{array}{rll}
	(1) & A    \to B  	& : A,\{\!|N_A|\!\}_{K_{AB}} \\
	(2) & B    \to A  	& : \{\!|N_A+1,N_B|\!\}_{K_{AB}}\\
	(3) & A    \to E(B) & : \{\!|N_B+1|\!\}_{K_{AB}}\\
	(4) & E(B) \to A    & : \{\!|N^{old}_A+1,N^{old}_B|\!\}_{K_{AB}}\\
\end{array}
$
&
$
\begin{array}{rll}
	(1\phantom{_1}) & A    \to B    & : A,\{\!|N_A|\!\}_{K_{AB}} \\
	(2\phantom{_1}) & B    \to A    & : \{\!|N_A+1,N_B|\!\}_{K_{AB}}\\
	(3\phantom{_1}) & A    \to E(B) & : \{\!|N_B+1|\!\}_{K_{AB}}\\
	(4\phantom{_1}) & E(B) \to G(A) & : \{\!|N_A+1,N_B|\!\}_{K_{AB}}\\
	(4_1) 		    & G(A) \to A    & : M_\mathit{fake}\\
	(4_1) & \multicolumn{2}{l}{A\text{ aborts}}
%	\multicolumn{3}{l}{A\text{ aborts: }sk(M_{fake}) \neq sk(\text{message } 4)} \\	
\end{array}
$
\\\\
\bottomrule
\end{tabular}
}
\end{center}
\end{table}

\subsection{Otway-Rees Protocol}

The Otway-Rees protocol (Table~\ref{Ot-ReProtocol} and~\cite{boydMathuria}) can be attacked through a 
type flaw attack in the last message. The guardian can defend $A$ if 
%and only if 
the considered scenario is the base case in Fig.~\ref{fig:casesE} and~\ref{fig:casesC} (both the configurations are possible since it is not mandatory that the information that the guardian can gain from the server $S$ have to be genuine), setting 
up two spy filters on $A$: (i) an outflow-spy filter in order to record in his 
dataset all of $A$'s outgoing messages that match the form of message (1) 
(removing agent names from the unencrypted part), and (ii) an inflow-spy 
filter in order to control whether amongst all of $A$'s incoming messages there 
is some message that is already in $D_G$. 
The interference consists in modifying the critical message in transit with a 
random message ``$I,M_\mathit{fake}$''. 
At the end of the protocol run, $A$ is forced to abort %because the decryption of $M_{fake}$ does not contain any critical data 
whereas $E$ thinks of having attacked the protocol. 
For this attack, the only possible situation that leads to a false positive is 
that in which the trusted third party generates a random key $K_{AB}$ such 
that $K_{AB}=\{I,A,B\}$, whereas false negatives are not possible since the 
attacker  knows neither the symmetric key shared between $A$ and $S$ 
nor that between $B$ and $S$; the only way to attack the protocol (sending the last 
message encrypted with that key) is to replay message (1) (its freshness is 
guaranteed by the nonce $N_A$). Instead, if the considered scenario is the base case in Fig.~\ref{fig:casesD} or Fig.~\ref{fig:casesF} $G$ can defend $A$ iff $E$ attacks the protocol with attack trace 2.

\begin{table}[t]
\caption{Otway-Rees Protocol.\label{Ot-ReProtocol}}
\begin{center}
\scalebox{0.8}{
\begin{tabular}{cc}
 \toprule
\multicolumn{2}{c}{{\bf Protocol}} \\
\midrule
\\
\multicolumn{2}{c}{$
\begin{array}{rll}
	(1) & A \to B & : I,A,B,\{\!|N_A,I,A,B|\!\}_{K_{AS}} \\
	(2) & B \to S & : I,A,B,\{\!|N_A,I,A,B|\!\}_{K_{AS}},\{\!|N_B,I,A,B|\!\}_{K_{BS}} \\
	(3) & S \to B & : I,\{\!|N_A,K_{AB}|\!\}_{K_{AS}},\{\!|N_B,K_{AB}|\!\}_{K_{BS}} \\
	(4) & B \to A & : I,\{\!|N_A,K_{AB}|\!\}_{K_{AS}} \\
\end{array}
 $}
\\\\
\toprule
{\bf Attack trace 1} & {\bf Interference 1}\\
\midrule 
\\
$
\begin{array}{rll}
	(1) & A    \to E(B) & :  I,A,B,\{\!|N_A,I,A,B|\!\}_{K_{AS}} \\
	(2) & B    \to S  	& : \text{Omitted} \\
	(3) & S    \to B  	& : \text{Omitted} \\
	(4) & E(B) \to A    & : I,\{\!|N_A,I,A,B|\!\}_{K_{AS}} \\
\end{array}
$
&
$
\begin{array}{rll}
	(1\phantom{_1}) & A    \to E(B) & : I,A,B,\{\!|N_A,I,A,B|\!\}_{K_{AS}} \\
	(2\phantom{_1}) & B    \to S    & : \text{Omitted} \\
	(3\phantom{_1}) & S    \to B    & : \text{Omitted} \\
	(4\phantom{_1}) & E(B) \to G(A) & : I,\{\!|N_A,I,A,B|\!\}_{K_{AS}} \\
	(4_1) 		    & G(A) \to A    & : I,M_\mathit{fake} \\
	(4_1) & {A\text{ aborts}}
	%D_{K_{AS}}(M_{fake}) \text{ does not contain }N_A}
\end{array}
$
\\\\
\toprule
\multicolumn{2}{c}{{\bf Attack trace 2}}\\
\midrule 
\\
\multicolumn{2}{c}{$
\begin{array}{rll}
	(1) & A    \to B  	& : I,A,B,\{\!|N_A,I,A,B|\!\}_{K_{AS}} \\
	(2) & B    \to E(S) & : I,A,B,\{\!|N_A,I,A,B|\!\}_{K_{AS}},\{\!|N_B,I,A,B|\!\}_{K_{BS}} \\
	(3) & E(S) \to B 	& : I,\{\!|N_A,I,A,B|\!\}_{K_{AS}},\{\!|N_B,I,A,B|\!\}_{K_{BS}} \\
	(4) & B    \to A 	& : I,\{\!|N_A,I,A,B|\!\}_{K_{AS}} \\
\end{array}
$}
\\\\
\toprule
\multicolumn{2}{c}{{\bf Interference 2}}\\
\midrule 
\\
\multicolumn{2}{c}{$
\begin{array}{rll}
	(1\phantom{_1}) & A    \to B	& : I,A,B,\{\!|N_A,I,A,B|\!\}_{K_{AS}} \\
	(2\phantom{_1}) & B    \to E(S) & : I,A,B,\{\!|N_A,I,A,B|\!\}_{K_{AS}},\{\!|N_B,I,A,B|\!\}_{K_{BS}} \\
	(3\phantom{_1}) & E(S) \to B    & : I,\{\!|N_A,I,A,B|\!\}_{K_{AS}},\{\!|N_B,I,A,B|\!\}_{K_{BS}} \\
	(4\phantom{_1}) & B \to G(A) & : I,\{\!|N_A,I,A,B|\!\}_{K_{AS}} \\
	(4_1) 		    & G(A) \to A    & : I,M_\mathit{fake} \\
	(4_1) & {A\text{ aborts}}%D_{K_{AS}}(M_{fake}) \text{ does not contain }N_A}
\end{array}
$}
\\\\
\bottomrule
\end{tabular}
}
\end{center}
\end{table}

\subsection{Encrypted Key Exchange Protocol}

The Encrypted-Key-Exchange protocol (Table~\ref{EKE} and~\cite{boydMathuria}) 
can be attacked by $E$ through a parallel sessions attack. The guardian can 
defend $A$ if the considered scenario is the base case in 
Fig.~\ref{fig:casesA}, once again, setting up two spy filters on $A$: (i) an 
outflow-spy filter in order to record in his dataset all of $A$'s outgoing 
messages which match the form of message $(3)$, and (ii) an inflow-spy filter 
in order to control whether amongst all of $A$'s incoming messages there is 
some message that is already in $D_G$. 
The interference consists in modifying the critical message in transit (message $(2.3_1)$) with a random message ($M_\mathit{fake}$).
This implies that $A$ generates for $B$ (in the parallel session) the correct response $\{D_R(M_\mathit{fake}),N_A'\}_R$ (message $(4)$, where $D_R(\cdot)$ is the symmetric decryption function), which is incorrect in the main session. 
%At the end of the protocol run $A$ aborts the main session because $N_{fake} \neq N_A$. 
For this attack, the only possible situation that leads to a false positive is that in which some agent $B$ starts the protocol with $A$ generating in message (3) an encrypted message $\{N_B\}_R$ such that it is already in $D_G$. 
Even though $D_G$ grows over time, if $R$ is large enough this probability 
remains negligible. 
False negatives are not possible; this is due to the fact that in order to attack the protocol, $E$ must reflect some message in the parallel session, but this behavior implies that the reflected message transited before through $G$ and thus the message belongs to $D_G$, so that $G$ can detect and defeat the ongoing attack. A guardian in the configuration of Fig.~\ref{fig:casesB} would not work, because $B$'s participation is not mandatory to attack the protocol and thus $E$ can easily exclude $G$ from the run of the protocol.

\begin{table}[t]
\caption{Encrypted Key Exchange Protocol.\label{EKE}}
\begin{center}
\scalebox{1.0}{
\begin{tabular}{cc}
\toprule
\multicolumn{2}{c}{{\bf Protocol}} \\
\midrule 
\\
\multicolumn{2}{c}{
 $
\begin{array}{rll}
	(1) & A \to B & : \{\!|K_A|\!\}_P \\
	(2) & B \to A & : \{\!|\{R\}_{K_A}|\!\}_P\\
	(3) & A \to B & : \{\!|N_A|\!\}_R \\
	(4) & B \to A & : \{\!|N_A,N_B|\!\}_R\\
	(5) & A \to B & : \{\!|N_B|\!\}_R\\
\end{array}
$}
\\\\
\toprule
{\bf Classical Attack} & {\bf Interference}\\
\midrule
\\
$
\begin{array}{rll}
	(1.1) & A 	 \to E(B) & : \{\!|K_A|\!\}_P \\
	(2.1) & E(B) \to A    & : \{\!|K_A|\!\}_P \\
	(2.2) & A 	 \to E(B) & : \{\!|\{R\}_{K_A}|\!\}_P \\
	(1.2) & E(B) \to A    & : \{\!|\{R\}_{K_A}|\!\}_P \\
	(1.3) & A 	 \to E(B) & : \{\!|N_A|\!\}_R \\
	(2.3) & E(B) \to A    & : \{\!|N_A|\!\}_R \\
	(2.4) & A 	 \to E(B) & : \{\!|N_A,N_B|\!\}_R \\
	(1.4) & E(B) \to A    & : \{\!|N_A,N_B|\!\}_R \\
	(1.5) & A 	 \to E(B) & : \{\!|N_B|\!\}_R \\
	(2.5) & E(B) \to A    & : \{\!|N_B|\!\}_R \\
\end{array}
$
&
$
\begin{array}{rll}
	(1.1\phantom{_1}) & A 	 \to E(B) & : \{\!|K_A|\!\}_P \\
	(2.1\phantom{_1}) & E(B) \to A    & : \{\!|K_A|\!\}_P \\
	(2.2\phantom{_1}) & A 	 \to E(B) & : \{\!|\{R\}_{K_A}|\!\}_P \\
	(1.2\phantom{_1}) & E(B) \to A    & : \{\!|\{R\}_{K_A}|\!\}_P \\
	(1.3\phantom{_1}) & A 	 \to E(B) & : \{\!|N_A|\!\}_R \\
	(2.3\phantom{_1}) & E(B) \to G(A) & : \{\!|N_A|\!\}_R \\
	(2.3_1) 	 	  & G(A) \to A    & : M_\mathit{fake} \\
	(2.4\phantom{_1}) & A 	 \to E(B) & : \{\!|N_\mathit{fake},N_B|\!\}_R \\
	(1.4\phantom{_1}) & E(B) \to A    & : \{\!|N_\mathit{fake},N_B|\!\}_R \\
	(1.5\phantom{_1}) & \multicolumn{2}{l}{A \text{ aborts}}% N_A \neq N_{fake}} \\
\end{array}
$
\\\\
\bottomrule
\end{tabular}
}
\end{center}
\end{table}

\subsection{SPLICE/AS Protocol}

The SPLICE/AS protocol (Table~\ref{SPLICE} and~\cite{Clark97:asurvey}) can be attacked by $E$ inducing $B$ to generate a correct answer for a message generated by $A$. 
The guardian can defend $A$ if %and only if 
the considered scenario is the base case in Fig.~\ref{fig:casesC}. 
When $AS$ receives the first request of the form $X,B,N_1$, the guardian can check, within the allowed temporal window, if the message sent to $AS$ in message (4) is of the form $B,Y,N_3$ where $Y \neq X$. 
The interference consists in stopping message (4) and making $A$ abort the protocol (again, using the ad-hoc flag in $A$).
For this attack, the only possible situation that leads to a false positive is when some agent $X$ starts the protocol with $B$ and also with $C$ (this implies that $S$ receives a message of the form $B,C,N_3$ at step (4)); obviously, since $C \neq A$ this situation is detected as an ongoing attack. 
False negatives are also possible if some agent $A$ starts the protocol with $B$ and also starts the protocol with $C$ near the “expiration” of the allowed temporal window; when $C$ responds, $G$'s temporal window has already “moved forward” and thus no flag will be raised. 
Another improbable situation for a false negative is when $G$ generates $N_\mathit{fake}$ such that $N_2 + 1 = N_\mathit{fake}$, but this probability remains negligible. A guardian in the configuration of the base case in Fig.~\ref{fig:casesD} could make $A$ abort too (raising the ad-hoc flag). Instead, the configurations in the base cases in Fig.~\ref{fig:casesE} e Fig.~\ref{fig:casesF} do not, obviously, work because not seeing $S$'s response $G$ cannot detect the attack.

\begin{table}[t]
\caption{SPLICE/AS Protocol.\label{SPLICE}}
\begin{center}
\scalebox{0.7}{
\begin{tabular}{cc}
 \toprule
{\bf Protocol} & {\bf Classical Attack} \\
\midrule
\\
$
\begin{array}{rll}
	(1) & A  \to AS & : A,B,N_1 \\
	(2) & AS \to A  & : AS,\{ AS,A,N_1,B,K_B \}_{K^{-1}_{AS}}\\
	(3) & A  \to B  & : A,B,\{A,T,L,\{N_2\}_{K_B}\}_{K^{-1}_A}\\
	(4) & B  \to AS & : B,A,N_3\\
	(5) & AS \to B  & : AS,\{ AS,B,N_3,A,K_A \}_{K^{-1}_{AS}}\\
	(6) & B  \to A  & : B,A,\{B,N_2+1\}_{K_A}\\
\end{array}
 $
&
$
\begin{array}{rll}
	(1\phantom{_1}) & A  \to AS   & : A,B,N_1 \\
	(2\phantom{_1}) & AS \to A    & : AS,\{ AS,A,N_1,B,K_B \}_{K^{-1}_{AS}}\\
	(3\phantom{_1}) & A  \to E(B) & : A,B,\{A,T,L,\{N_2\}_{K_B}\}_{K^{-1}_A}\\
	(3_1) 		    & E  \to B    & : E,B,\{E,T,L,\{N_2\}_{K_B}\}_{K^{-1}_E}\\
	(4\phantom{_1}) & B  \to AS   & : B,E,N_3\\
	(5\phantom{_1}) & AS \to B    & : AS,\{  AS,B,N_3,E,K_E \}_{K^{-1}_{AS}}\\
	(6\phantom{_1}) & B  \to E    & : B,E,\{B,N_2+1\}_{K_A}\\
	(6_1) 		    & E  \to A    & : B,A,\{B,N_2+1\}_{K_A}\\
\end{array}
$
 
\\\\
\toprule
 \multicolumn{2}{c}{{\bf Interference}}\\
\midrule 
\\
\multicolumn{2}{c}{$
\begin{array}{rll}
(1\phantom{_1}) & A     \to AS    & : A,B,N_1 \\
(2\phantom{_1}) & AS    \to A     & : AS,\{ AS,A,N_1,B,K_B \}_{K^{-1}_{AS}}\\
(3\phantom{_1}) & A     \to E(B)  & : A,B,\{A,T,L,\{N_2\}_{K_B}\}_{K^{-1}_A}\\
(3_1) 		    & E     \to B     & : E,B,\{E,T,L,\{N_2\}_{K_B}\}_{K^{-1}_E}\\
(4\phantom{_1}) & B     \to G(AS) & : B,E,N_3\\
(5\phantom{_1}) & G(E)  \to A     & : B,A,\{B,N_\mathit{fake}\}_{K_A}\\
(6\phantom{_1}) & G(AS) \to AS    & : B,E,N_3\\
(6\phantom{_1}) & {A\text{ aborts}}%N_2 + 1 \neq N_{fake}}
\end{array}
$}
\\\\
\bottomrule
\end{tabular}
}
\end{center}
\end{table}

\subsection{Boyd-Mathuria Example (fixed for masquerading attack)}

We have modified the BME protocol (Table~\ref{BMEfix}, the original version can be found in~\cite{boydMathuria}) adding agent names in the encrypted segments of the message (2) in order to avoid the masquerading attack given in~\cite{boydMathuria}. 
However, since messages do not have a temporal collocation (timestamps are not present) and nonces are not used, the protocol is still vulnerable to a replay attack. 
The guardian can defend $B$ if %and only if 
the considered scenario is the base case in Fig.~\ref{fig:casesD} setting up 
an inflow-spy filter and an outflow-spy filter on $S$ and only an inflow-spy filter on $B$. 
When $S$ receives the first request of the form $X,B$, the guardian can check if the message received by $B$, within the allowed temporal window, has been just 
generated by $S$ (i.e. if $m$ belongs to those messages added to $D_G$ in the last $j$-steps of the dataset evolution).
The interference simply consists in stopping the last message of the protocol.
False positives are possible if some agent starts the protocol with $B$ near 
the $G$'s temporal window expiration, whereas false negatives are not possible 
since $G$ continues to move forward his temporal window. 
However, attack traces 2 and 4 do not defend $A$ from the attack but only $B$. 
Moreover, a guardian in the configuration of the base case in 
Fig.~\ref{fig:casesC} could defend $A$ only for attack traces 2 and 4. 
Instead, $G$ in the configuration of the base cases in Fig.~\ref{fig:casesE} 
and Fig.~\ref{fig:casesE} does not work, because not seeing $S$'s response he 
cannot detect the attack.

\begin{table}[t]
\caption{Boyd-Mathuria Example (Fixed for masquerading attack).\label{BMEfix}}
\begin{center}
\scalebox{0.7}{
\begin{tabular}{cc}
\toprule
\multicolumn{2}{c}{{\bf Protocol}}\\
\midrule
\\
\multicolumn{2}{c}{
$
\begin{array}{rll}
	(1) & A \to S & : A,B \\
	(2) & S \to A & : \{\!|K_{AB},B|\!\}_{K_{AS}},\{\!|K_{AB},A|\!\}_{K_{BS}}\\
	(3) & A \to B & : \{\!|K_{AB},A|\!\}_{K_{BS}}\\
\end{array}
$
}
\\\\
\toprule
{\bf Attack trace 1} & {\bf Interference 1} \\
\midrule 
\\
$
\begin{array}{rll}
	(1) & A    \to S & : \text{Omitted} \\
	(2) & S    \to A & : \text{Omitted} \\
	(3) & E(A) \to B & : \{\!|K'_{AB},A|\!\}_{K_{BS}}\\
\end{array}
$
&
$
\begin{array}{rll}
	(1\phantom{_1}) & A \to S       & : \text{Omitted} \\
	(2\phantom{_1}) & S \to A       & : \text{Omitted} \\
	(3\phantom{_1}) & E(A) \to G(B) & : \{\!|K'_{AB},A|\!\}_{K_{BS}}\\
	(3_1) 			& \multicolumn{2}{l}{G\text{ stops message }3} \\
	
\end{array}
$
\\\\
\toprule
\multicolumn{2}{c}{{\bf Attack trace 2}} \\
\midrule 
\\
\multicolumn{2}{c}{$
\begin{array}{rll}
	(1\phantom{_1}) & A    \to S    & : A,B \\
	(2\phantom{_1}) & S    \to E(A) & : \{\!|K_{AB},B|\!\}_{K_{AS}},\{\!|K_{AB},A|\!\}_{K_{BS}}\\
	(2_1) 		    & E(A) \to A    & : \{\!|K'_{AB},B|\!\}_{K_{AS}},\{\!|K'_{AB},A|\!\}_{K_{BS}}\\
	(3\phantom{_1}) & A    \to B    & : \{\!|K'_{AB},A|\!\}_{K_{BS}}\\
\end{array}
$}
\\\\
\toprule
\multicolumn{2}{c}{{\bf Interference 2}}\\
\midrule 
\\
\multicolumn{2}{c}{$
\begin{array}{rll}
	(1\phantom{_1}) & A    \to S    & : A,B \\
	(2\phantom{_1}) & S    \to E(A) & : \{\!|K_{AB},B|\!\}_{K_{AS}},\{\!|K_{AB},A|\!\}_{K_{BS}}\\
	(2_1) 		    & E(A) \to A    & : \{\!|K'_{AB},B|\!\}_{K_{AS}},\{\!|K'_{AB},A|\!\}_{K_{BS}}\\
	(3\phantom{_1}) & A    \to G(B) & : \{\!|K'_{AB},A|\!\}_{K_{BS}}\\
	(3_1) 			& \multicolumn{2}{l}{G\text{ stops message }3} \\
\end{array}
$}
\\\\
\toprule
\multicolumn{2}{c}{{\bf Attack trace 3}}\\
\midrule 
\\
\multicolumn{2}{c}{$
\begin{array}{rll}
	(1) & E(A) \to S    & : A,B \\
	(2) & S    \to E(A) & : \{\!|K_{AB},B|\!\}_{K_{AS}},\{\!|K_{AB},A|\!\}_{K_{BS}}\\
	(3) & E(A) \to B    & : \{\!|K'_{AB},A|\!\}_{K_{BS}}\\
\end{array}
$}
\\\\
\toprule
\multicolumn{2}{c}{{\bf Interference 3}} \\
\midrule
\\
\multicolumn{2}{c}{$
\begin{array}{rll}
	(1\phantom{_1}) & E(A) \to S    & : A,B \\
	(2\phantom{_1}) & S    \to E(A) & : \{\!|K_{AB},B|\!\}_{K_{AS}},\{\!|K_{AB},A|\!\}_{K_{BS}}\\
	(3\phantom{_1}) & E(A) \to G(B) & : \{\!|K'_{AB},A|\!\}_{K_{BS}}\\
	(3_1) 			& \multicolumn{2}{l}{G\text{ stops message }3} \\
\end{array}
$}
\\\\
\toprule
\multicolumn{2}{c}{{\bf Attack trace 4}} \\
\midrule 
\\
\multicolumn{2}{c}{$
\begin{array}{rll}
	(1) & A    \to E(S) & : A,B \\
	(2) & E(S) \to A    & : \{\!|K'_{AB},B|\!\}_{K_{AS}},\{\!|K'_{AB},A|\!\}_{K_{BS}}\\
	(3) & A    \to B    & : \{\!|K'_{AB},A|\!\}_{K_{BS}}\\
\end{array}
$}
\\\\
\toprule
\multicolumn{2}{c}{{\bf Interference 4}} \\
\midrule
\\
\multicolumn{2}{c}{$
\begin{array}{rll}
	(1\phantom{_1}) & A    \to E(S)  & : A,B \\
	(2\phantom{_1}) & E(S) \to A     & : \{\!|K'_{AB},B|\!\}_{K_{AS}},\{\!|K'_{AB},A|\!\}_{K_{BS}}\\
	(3\phantom{_1}) & A    \to G(B)  & : \{\!|K'_{AB},A|\!\}_{K_{BS}}\\
	(3_1) 			& \multicolumn{2}{l}{G\text{ stops message }3} \\
\end{array}
$}
\\\\
\bottomrule
\end{tabular}
}
\end{center}
\end{table}

\end{LONG}

\end{document}